\PassOptionsToPackage{dvipsnames}{xcolor}
\documentclass[sigconf]{acmart}
\usepackage{popets}
\setcopyright{popets}
\copyrightyear{2026}
\acmYear{2026}
\acmVolume{YYYY}
\acmNumber{X}
\acmDOI{XXXXXXX.XXXXXXX}
\acmISBN{}
\acmConference{ACM conference}
\usepackage[T1]{fontenc}
\usepackage{amsmath,amsfonts}
\usepackage{amsthm}
\usepackage{graphicx}
\usepackage{booktabs}
\usepackage{multirow}
\usepackage{array}
\usepackage{adjustbox}
\usepackage{xspace}
\usepackage{url}
\usepackage{xcolor}
\usepackage{tikz}
\usetikzlibrary{arrows.meta,positioning}
\usepackage{microtype}
\definecolor{kred}{HTML}{FF0052}
\definecolor{kyellow}{HTML}{FFD400}
\definecolor{kgreen}{HTML}{00C688}
\definecolor{kblue}{HTML}{0055DA}
\graphicspath{{figures/}}
\def\name{KeHoSu\xspace}

\theoremstyle{plain}
\newtheorem{proposition}{Proposition}
\newtheorem{corollary}[proposition]{Corollary}
\theoremstyle{definition}
\newtheorem{definition}{Definition}

\begin{document}

\title{KeHoSu: \textbf{Ke}yed \textbf{Ho}moglyph \textbf{Su}bstitution for Authenticated Text Provenance over Copy-Paste Channels}

\author{Andrei Alexandru Dr{\'a}ghici}
\affiliation{%
  \institution{University of Bucharest}
  \city{Bucharest}
  \country{Romania}
  }
\email{andrei-alexandru.draghici@s.unibuc.ro}

\author{Sediola Ruko}
\affiliation{%
  \institution{Sapienza University of Rome}
  \city{Rome}
  \country{Italy}
  }
\email{ruko@di.uniroma1.it}

\author{Briland Hitaj}
\affiliation{%
  \institution{Confidencial Inc}
  \state{Menlo Park, CA}
  \country{USA}
}
\email{briland.hitaj@confidencial.io}

\author{Dorjan Hitaj}
\affiliation{%
 \institution{Sapienza University of Rome}
 \city{Rome}
 \country{Italy}
 }
\email{hitaj.d@di.uniroma1.it}

\renewcommand{\shortauthors}{Dr{\'a}ghici et al.}

\begin{abstract}
Websites, publishing platforms, and, more recently, large language models produce and distribute text on a large scale. Much of this material is redistributed via "copy-paste", which strips away file metadata but preserves Unicode code points. Many common letters have counterparts in other scripts that appear identical, despite having distinct code points, and are rendered as the same glyph by most fonts.

In this paper, we introduce \textbf{\name}, a keyed attribution method that embeds an authenticated provenance identifier by leveraging the choice between visually confusable Unicode code point variants. \name selects between a Latin letter and its visually identical Cyrillic counterpart to encode one bit of an encrypted and authenticated identifier of the platform that produced the text and the specific user account on that platform. \name employs rateless codes to spread the bits, allowing it to keep no positional state.
We demonstrate that \name is invariant to the reordering of sentences (up to a constant margin) and that, in the case of partial excerpts, it depends solely on their length. Furthermore, we define the limits regarding the capacity, removability, and detectability of this code-point-based channel. 
In our evaluation \name survived "copy-paste" operations across $18$ different platforms. Moreover, \name achieves $50\%$ recovery relying on approximately $173$ letters on average, typically corresponding to one or two sentences.

\end{abstract}


\maketitle

\section{Introduction} \label{sec:intro}
Large language models (LLMs) make fluent text cheap to produce at scale, and influence operations already exploit this capability~\cite{goldstein, ferrara}. A botnet of over $1{,}000$ Twitter accounts, for instance, posted ChatGPT-generated content that state-of-the-art classifiers could not tell apart from human accounts~\cite{aibotnet}. When a suspicious comment, post, or article turns up, a moderator or investigator wants to know which platform generated it and for which account, and regulatory bodies increasingly ask providers to mark their output in a machine-readable format~\cite {euaiact}. For files, provenance metadata such as C2PA content credentials~\cite{c2pa} answers this question, although even there metadata is often stripped when content is shared~\cite{nistsynth}. Text, however, rarely travels as a file. It is quoted, re-posted, and forwarded by copy-paste, which drops file-level metadata and, depending on the destination, formatting, but preserves the Unicode code points.
Existing tools do not attribute text that travels this way. Detectors of machine-generated text try to estimate whether a model wrote a text, but they are easily evaded by paraphrasing~\cite{krishna,sadasivan}, and, in cases when they work, they are unable to say anything about who received the text. Generation-time watermarks~\cite{kirchenbauer,synthidtext} bias the token distribution of one model and detect machine origin once enough tokens are observed. However, they carry at most a few bits and require control over the model's sampling, so a platform cannot apply them to text it did not generate. Similarly, multi-bit variants~\cite{mpac,credid,qu} embed a user identifier in the same distribution. On the other hand, format-based marks hide data in Unicode spaces and zero-width characters~\cite{unispach,aitsteg}, which input filters can strip~\cite{zew}. Another direction is watermarking text by substituting homoglyphs, i.e., letters of other scripts that look like Latin letters~\cite {rizzo}. Many common letters have counterparts in other scripts that look the same. For example, the Latin \texttt{a} (U+0061) and the Cyrillic \texttt{a} (U+0430) are distinct code points, but most fonts draw using the same glyph.

In this paper, we introduce \name, which embeds an authenticated provenance identifier within a selection of visually confusable variants. The identifier consists of a platform field and a user field. This identifier is encrypted and authenticated, expanded using a rateless code, and embedded into the carriers whose local context, when processed via a keyed hash function, corresponds to the index of each encoded bit. On the other end, the extractor recomputes the hashes from the received text, requiring neither the original document nor the specific location from which the extracted fragment originated. Since the mark is applied to finished text, the \name watermark does not depend on how the text was produced. A website or a publishing platform can mark the copy it serves to each account, and an organization can mark each copy of an internal document, so that a quoted or leaked excerpt can be traced to its recipient. 

In this paper, we make the following contributions:
\begin{itemize}
\item We design \name, a keyed, authenticated, multi-bit text watermark carried by code points. Carriers are indexed by their sentence-clipped local content, so that decoding resynchronizes after insertions and deletions.

\item We analyze the code-point channel and determine the limits for the recoverable payload and the minimum markable length. We demonstrate that \name is unaffected by the reordering of sentences (up to a constant margin) and that, in the case of partial excerpts, it depends solely on their length. 

\item We also present an alternative method in which anyone with the public key can verify the originating platform, while the user's identity remains encrypted. Only the platform can decrypt both the platform identifier and the user identifier.

\item We evaluate \name using prose texts taken from public-domain books and stories generated by LLMs. \name copy-paste operations across 18 platforms, and the $50\%$ recovery threshold for a $72$-bit identifier lies at $167$ received characters for texts from books, and at $178$ characters for texts generated by LLMs.

\end{itemize}

\section{Background} \label{sec:background}
In this section we briefly introduce the Unicode mechanisms and the coding and cryptographic primitives that \name builds upon.
\subsection{Confusable characters} \label{sec:bg:confusables}
Unicode assigns distinct code points to letters of different scripts even when they share a glyph. The Latin \texttt{a} (U+0061) and the Cyrillic \texttt{a} (U+0430) are such a pair: most fonts draw them identically, and, as a consequence, most readers cannot tell which one a text comprises. Unicode Technical Standard \#39~\cite{tr39} lists these pairs as confusable and introduces two methods for handling them, namely the skeleton and the mixed-script detection. The skeleton maps every character to a representative of its confusable class. Mixed-script detection flags a word whose letters come from more than one script. Yet, confusables continue to be used as an attack vector. Homograph attacks have been used to register domain names that resemble trusted ones~\cite {homograph}. On the other hand, invisible and confusable code points hide logic in source code~\cite{trojansource} and perturb the input of natural language processing (NLP) models~\cite{badchars}, and homoglyph substitution was shown to evade detectors of machine-generated text~\cite{silverspeak}.

\subsection{Rateless codes} \label{sec:bg:codes}
A rateless (fountain) code expands $K$ source bits into an unbounded stream of coded symbols, each a function of a subset of the source bits, such that any $(1+\varepsilon)K$ received symbols recover the source with high probability, where $\varepsilon$ is the overhead of the rateless code~\cite{ltcodes,raptor}. 
In a random linear fountain over $\mathrm{GF}(2)$, each coded bit is the parity of a random subset of source bits. Decoding is done via Gaussian elimination. The expected excess over $K$ for a random binary matrix to reach full rank is about $1.6$ symbols, independent of $K$~\cite{mackay}. 

\subsection{Authenticated encryption and signatures} \label{sec:bg:crypto}
An authenticated encryption scheme protects both the confidentiality and the integrity of a message. Its ciphertext-integrity advantage bounds the probability that an adversary without the key produces a new ciphertext that decrypts~\cite{etm}. Deterministic authenticated encryption, as in the SIV mode~\cite{siv}, derives the initialization vector from the message itself and needs no transmitted nonce. AES-GCM-SIV~\cite{rfc8452} is a nonce-misuse-resistant instance of this design. HMAC~\cite{hmac} is the standard keyed hash for message authentication. BLAKE2~\cite{blake2} is a fast keyed hash that we use as a pseudorandom function. A signature scheme secure against existential forgery under adaptive chosen-message attacks (EUF-CMA)~\cite{gmr}, such as EdDSA~\cite{eddsa}, lets any holder of the public key verify a message signed with the secret key.

\section{Threat Model} \label{sec:threat}
We consider services that release text: a website, a publishing platform or an LLM platform that marks every text it returns to its users. The text then travels over a redistribution channel. It is selected, copied and pasted elsewhere, as a quote, a repost or a forwarded message. A party that finds a suspicious sample submits it for attribution and wants to learn the platform and the account the text was released to.

The scheme is a pair of keyed algorithms. The embedder $E_k(T,m)\!\to\!T'$ chooses carrier variants that encode $m$ and keeps $\sigma(T')=\sigma(T)$, where T is the text, m the identifier and $T'$ the marked text. The extractor $D_k(\tilde T)$ returns an identifier or $\bot$. The reader distortion $d(T,T')=|\{i:\text{$T$ and $T'$ render differently at $i$}\}|$ measures perceptible change. When every variant set renders as one glyph, $d(T,E_k(T,m))=0$; a pair that some font draws differently is removed from the carrier set.
The received text is $\tilde T=\mathcal{A}(T')$, where the channel
\begin{equation}
\mathcal{A}=\eta\circ\mathcal{S}\circ\pi
\label{eq:channel}
\end{equation}
models the resharer. The copy-paste transport $\pi$ preserves code points but may drop formatting and metadata. The structural stage $\mathcal{S}$ retains a subset of the sentences in an arbitrary order. The edit stage $\eta$ erases or corrupts surviving carriers, at a total rate $p_e$. $\mathcal{S}$ removes whole spans, while $\eta$ acts on single characters.

Our threat model includes the following actors:\\
\noindent \textbf{Embedder.} The platform holds a secret key $k$. Every text that leaves the platform carries the identifier $m=(\mathrm{pid},\mathrm{uid})$ of the platform and of the user account, written by the embedder after the text has been produced.\\
\noindent\textbf{Extractor.} Any party holding $k$, typically the platform or an investigator it authorizes. It receives a sample and outputs either $(\mathrm{pid},\mathrm{uid})$ or $\bot$ (unmarked or invalid). \\
\noindent\textbf{Resharer.} It has no intent to remove or forge the mark. It copies and pastes, and may keep only some of the sentences, reorder them, split them across several posts, mix them with its own text and make light edits.\\
\noindent \textbf{Forger.} Its goal is to link a victim's identifier to text that the victim never received. The forger does not hold $k$ and cannot query the embedder for identifiers other than its own.  It can collect any amount of text marked for the victim and for itself, splice it into other text, unmarked or marked with another identifier, and submit up to $q$ samples to a verifier. 

\section{\name} \label{sec:method}
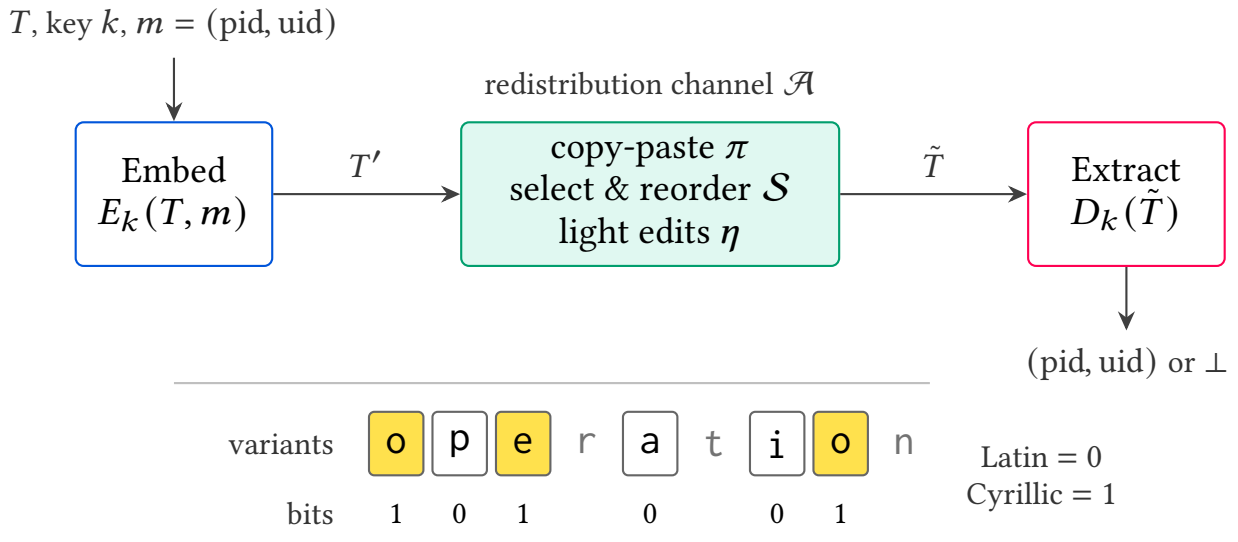
\begin{figure}[t]
\centering
\begin{tikzpicture}[
    font=\footnotesize, >={Stealth[length=3.6pt,width=3.2pt]},
    box/.style={draw=black!70, rounded corners=1.5pt, align=center, inner sep=2.5pt,
                minimum height=9.5mm},
    lab/.style={font=\scriptsize, text=black!80, align=center},
    car/.style={draw=black!60, rounded corners=1pt, minimum width=3.6mm, minimum height=4.2mm,
                inner sep=0pt, font=\ttfamily\small},
    cyr/.style={car, fill=kyellow!70},
    lat/.style={car, fill=white},
    non/.style={minimum width=3.6mm, minimum height=4.2mm, inner sep=0pt,
                font=\ttfamily\small, text=black!55},
  ]
  \node[box, draw=kblue, minimum width=13mm] (emb) at (0,0) {Embed\\$E_k(T,m)$};
  \node[box, fill=kgreen!12, draw=kgreen!80!black, minimum width=25mm] (ch) at (3.15,0)
       {copy-paste $\pi$\\select \& reorder $\mathcal{S}$\\light edits $\eta$};
  \node[box, draw=kred, minimum width=13mm] (ext) at (6.3,0) {Extract\\$D_k(\tilde T)$};
  \node[lab, above=0.6mm of ch] {redistribution channel $\mathcal{A}$};
  \draw[->, black!75] (emb) -- node[lab, above] {$T'$} (ch);
  \draw[->, black!75] (ch) -- node[lab, above] {$\tilde T$} (ext);
  \node[lab, above=4.2mm of emb] (in) {$T$, key $k$, $m=(\mathrm{pid},\mathrm{uid})$};
  \draw[->, black!75] (in) -- (emb);
  \node[lab, below=4.2mm of ext] (out) {$(\mathrm{pid},\mathrm{uid})$ or $\bot$};
  \draw[->, black!75] (ext) -- (out);
  \draw[black!25] (0,-1.25) -- (5,-1.25);
  \begin{scope}[shift={(1.47,-1.65)}]
    \node[lab, anchor=east] at (-0.3,0) {variants};
    \node[lab, anchor=east] at (-0.3,-0.46) {bits};
    \foreach \ch/\st/\b [count=\i from 0] in
        {o/cyr/1, p/lat/0, e/cyr/1, r/non/{}, a/lat/0, t/non/{}, i/lat/0, o/cyr/1, n/non/{}} {
      \node[\st] at (0.42*\i,0) {\ch};
      \node[font=\scriptsize] at (0.42*\i,-0.46) {\b};
    }
    \node[lab, anchor=west] at (3.66,-0.23) {Latin $=0$\\Cyrillic $=1$};
  \end{scope}
\end{tikzpicture}
\caption{Overview of \name. Carriers are letters with an identical-looking Cyrillic counterpart. The Latin variant encodes bit~$0$ and the Cyrillic variant bit~$1$. Each carrier holds the coded bit whose index is a keyed hash of its local context.}
\label{fig:system}
\end{figure}

Given a text $T$, a key $k$ and an identifier $m=(\mathrm{pid},\mathrm{uid})$, the embedder $E_k$:
\begin{enumerate}
\item seals $m$ into a $K$-bit payload with a truncated authentication tag;
\item expands the payload with a rateless code into a stream of coded bits indexed by $j\in\{0,\dots,\Lambda-1\}$; where $\Lambda$ corresponds to the number of indices addressed by the placement.
\item for every carrier $i$, computes the index $j_i$ as a keyed hash of the carrier's normalized local context and writes coded bit $j_i$ as the variant of $i$.
\end{enumerate}
The extractor $D_k$ reverses these steps. It reads the variant of every carrier as a vote for the coded bit at that carrier's index, solves the code, removes the mask and accepts only if the tag verifies. Neither side keeps any positional state, since the bit a carrier holds depends only on the text around it and on $k$.

\subsection{Carrier set} \label{sec:carriers}
A carrier pairs a Latin letter with a Cyrillic letter that renders as the same glyph. The Latin variant encodes bit~$0$ and the Cyrillic variant bit~$1$. As shown in Figure~\ref{fig:system}, the word \texttt{operation}, for example, holds six carriers (o, p, e, a, i, and o). A letter is admitted only if both of its cases have a Cyrillic homoglyph. Recasing then maps carriers to carriers, never creating an unmarked carrier that the decoder would read as data. 
The rule excludes the capitals B, H, K, M, and T, whose Cyrillic look-alikes (U+0412, 041D, 041A, 041C, 0422) have no Latin-looking lowercase. Every other character is emitted unchanged.

\subsection{Sealing the identifier} \label{sec:payload}
The identifier is encrypted and authenticated before coding, so that a recovered payload is self-validating and reveals nothing about $m$ to a party without $k$. A key-rotation counter selects the key $k_{\text{epoch}}$; the counter is not transmitted. To keep the payload short we use no random nonce and seal $m$ deterministically, in the style of SIV~\cite{siv} rather than the encrypt-then-MAC composition~\cite{etm}:
\begin{equation}
(\mathrm{ct},\tau)=\mathrm{SIV}_{k_{\text{epoch}}}\big(\mathrm{pid}\,\Vert\,\mathrm{uid}\big),\qquad
\text{payload}=\mathrm{ct}\,\Vert\,\tau_{1:t},
\label{eq:payload}
\end{equation}
where $\mathrm{ct}$ is the encrypted $(p{+}u)$-bit identifier and $\tau_{1:t}$ the tag truncated to $t$ bits (a truncated HMAC~\cite{hmac} tag serves equally well).
A SIV mode derives its keystream from the full tag; we derive it from the truncated tag, $\mathrm{ct}=m\oplus F_k(\tau_{1:t})$, so that the extractor can recompute it. A candidate $(\mathrm{ct}',\tau')$ is accepted when $\mathrm{MAC}_k\big(\mathrm{ct}'\oplus F_k(\tau')\big)_{1:t}=\tau'$. The construction is MAC-then-mask on the tag function of AES-GCM-SIV~\cite{rfc8452}. 
We instantiate it with AES-256-GCM-SIV under a fixed nonce: $\tau_{1:t}$ is the first $t=32$ bits of the GCM-SIV tag over a domain byte and $m$, and $F_k(\tau_{1:t})$ is the GCM-SIV encryption of the domain byte and $\tau_{1:t}$, truncated to $|m|$. Verification recomputes the seal from the recovered $m$ and compares it in constant time. Keys rotate before $2^{t/2}$ identifiers ($65{,}536$ at $t=32$) are sealed under one key, which bounds keystream collisions.

\subsubsection{Asymmetric mode} For verification without $k$, the platform holds a signature key pair $(\mathrm{sk},\mathrm{pk})$ and an encryption key $K_P$, and embeds
\begin{equation}
\text{payload}=\mathrm{pid}\,\Vert\,\mathrm{Sign}_{\mathrm{sk}}(\mathrm{pid}\,\Vert\,h)\,\Vert\,\mathrm{Enc}_{K_P}(\mathrm{uid}),
\label{eq:asym}
\end{equation}
where $h=H(\sigma(T))$ hashes the normalized content and $\mathrm{Enc}_{K_P}$ is the authenticated encryption of the symmetric mode. Anyone holding $\mathrm{pk}$ verifies the platform of origin, while only the platform reads the user identifier. A BLS or EdDSA~\cite{eddsa} signature takes $384$--$512$ bits and brings the payload to $456$--$584$ bits.
The two modes can run together as separate marks on disjoint carrier subsets.

\subsection{Content-anchored placement} \label{sec:placement}
A keyed hash of each carrier's local context, reduced modulo $\Lambda$, gives the index of the coded bit that the carrier holds:
\begin{equation}
j_i \;=\; H_k\big(\mathrm{ctx}(i)\big) \bmod \Lambda ,
\label{eq:index}
\end{equation}
where $\mathrm{ctx}(i)$ is the window of $W_{\mathrm{ctx}}$ characters centered on $i$, with every carrier in it folded to its skeleton and lower-cased. Normalization removes the variant choice from the context, so the index does not depend on the embedded bit, and the embedder and the decoder derive the same mapping without a circular dependence on the payload. At most $\Lambda$ distinct coded bits can be addressed, and $\Lambda$ must exceed $K$ with room for the occupancy loss measured in Section~\ref{sec:evalsetup}. Two carriers with the same context receive the same index, hence the same coded bit, and the decoder takes a majority vote over such copies. We write $\nu(L)\le L$ for the number of distinct indices among $L$ carriers.
The index depends only on local content. An insertion or deletion changes the indices of the carriers whose window it overlaps and of no others, whereas a hash of the whole text, or an absolute offset, would desynchronize every carrier after a single inserted character. The positional codec of earlier homoglyph marks~\cite{rizzo} lacks this property (see Section~\ref{sec:evalbaseline}).

\subsubsection{Span clipping} 
A window near a sentence boundary also covers part of the neighboring sentence, and would change whenever that neighbor is deleted or moved. We therefore cut the window at the boundaries of the carrier's own sentence. A carrier's index then depends only on its sentence and is unchanged when other sentences are deleted or reordered (see Proposition~\ref{prop:perm}). Two rules keep the clipped spans stable. First, the embedder and the decoder segment the skeleton $\sigma(\cdot)$, whose boundaries do not depend on the embedded bits. A segmenter that reads the marked characters would not recognize a sentence beginning with a Cyrillic capital. Second, a UAX\,\#29 segmenter~\cite{uax29} assigns the whitespace after a sentence to that sentence, and the clipped span excludes this separator. Without the second rule, the last carriers of a sentence would read one context inside a document and another at its end.

\subsection{Coding and extraction} \label{sec:ecc}
\name relies on a rateless fountain code~\cite{ltcodes,raptor,mackay} to protect the payload. This allows any $(1+\varepsilon)K$ correctly received coded bits at distinct indices to decode it, with a small overhead $\varepsilon$. Truncation and deletion erase coded bits, which the code tolerates. An edit inside a carrier's context window does more damage: it mis-indexes the carrier, whose bit is then counted at a wrong index. This removes a vote from the right index and adds one at a wrong one, so an edit produces an error rather than an erasure.

Given a sample $\tilde T$, the extractor proceeds in four steps.
\begin{enumerate}
    \item Locate the carriers by testing each character against the confusable set, and reads each variant as a bit.
    \item Recompute the index of each carrier from its context and $k$, skipping the carriers within $W_{\mathrm{ctx}}/2$ of either end of the sample, whose windows may be incomplete.
    \item Sum the votes per index and solves the code. The elimination decoder takes the $K+16$ equations with the largest vote margins, solves them by Gaussian elimination over $\mathrm{GF}(2)$, and redraws subsets up to eight times until a candidate passes the tag.
    \item Remove the mask and check the tag, returning $(\mathrm{pid},\mathrm{uid})$ on success and $\bot$ otherwise.
\end{enumerate}

Pure-Latin text reads $0$ at every carrier and decodes to the all-zero payload, which the tag rejects only after decoding, incurring verification queries.
Moreover, rejecting any sample that holds non-Latin carriers before step $3$ avoids this cost, as we show in Section~\ref{sec:evalfalse}.
The carriers skipped in step~$2$ form a boundary layer of about $W_{\mathrm{ctx}}\alpha_c$ carriers, where $\alpha_c$ is the number of carriers per character, spaces, and punctuation included. The embedder refuses a text unless its carriers outside the boundary layer address at least $K+8$ distinct indices, a margin that covers the spread of the code overhead around its mean.

\subsection{Error-tolerant decoding} \label{sec:decoders}
Elimination treats every equation as correct, so a single mis-indexed carrier on a chosen equation corrupts the solution. Near the recovery threshold there is little redundancy to outvote it. We consider two ways to tolerate such errors:

\begin{enumerate}
    \item \textbf{Consensus decoding.} The decoder first runs the elimination decoder. If no candidate verifies, it repeatedly draws $K+4$ equations, keeping the indices on which two or more carriers agree and filling the remainder at random. A solution is tag-checked only if it agrees with at least three quarters of the equations left out of the draw. The search stops after $4000$ draws or $64$ further tag checks, so a sample costs at most $8+64=72$ tag checks.
    \item \textbf{Outer code.} A Reed--Solomon $[15,9]$ code~\cite{reedsolomon} over $\mathrm{GF}(2^8)$ wraps the payload before the fountain code ($120$ coded bits, three symbol errors corrected). The outer decoder sees the payload only after elimination has produced a solution.
\end{enumerate}

\subsection{Channel analysis} \label{sec:theory}
We characterize the code-point channel by its capacity, its robustness to structural and character-level edits, its detectability and its security.  We refer the reader to Appendix~\ref{app:proofs} for the proofs that have been omitted from this section.

\subsubsection{Capacity} \label{sec:capacity}
Let a text contain $N$ letters, and let $\alpha$ be the fraction of them that are carriers. Summing the English letter frequencies~\cite{norvig} of the lower-case carriers gives $12.49\,(\text{e})+8.04\,(\text{a})+7.64\,(\text{o})+7.57\,(\text{i})+6.51\,(\text{s})+3.34\,(\text{c})+2.14\,(\text{p})+1.66\,(\text{y})+0.23\,(\text{x})+0.16\,(\text{j})=49.78\%$ of all letters. Capitals are a few percent of the letters in running prose and move this figure by less than one point. A $10\%$ discount for pairs whose glyphs differ in some fonts gives $\alpha\in[0.44,0.50]$, and we size the scheme with $\alpha=0.45$. A text of $N$ letters has $L\approx\alpha N$ carriers with raw content $R=\sum_{i\in\mathcal{C}}\log_2 v_i=bL$ bits.

\begin{proposition}[Recoverable payload]
\label{prop:capacity}
Fix a text with $L$ carriers of raw content $R=bL$ bits, and model the reshare as an erasure channel that loses each carrier with probability $p_e$, with arbitrary correlation among losses. Any extractor recovers at most the content of the surviving carriers,
\begin{equation}
K \;\le\; (1-p_e)\,R \;=\;(1-p_e)\,bL \;\approx\;(1-p_e)\,b\,\alpha N,
\label{eq:capacity}
\end{equation}
and a rateless code attains this within a factor $(1+\varepsilon)$. Writing $r=K/R$ for the payload per raw carrier bit, a target $K$ is infeasible if $r>1-p_e$ and feasible if $r\le (1-p_e)/(1+\varepsilon)$, i.e., if $p_e\le 1-(1+\varepsilon)r$.
\end{proposition}

\begin{corollary}[Minimum markable length]
\label{cor:nmin}
On the clean channel, where $K\approx b\,r\,\alpha N$, an identifier of $p+u+t$ bits requires
\begin{equation}
N_{\min}\;\approx\;\frac{p+u+t}{b\,r\,\alpha}
\label{eq:nmin}
\end{equation}
letters.
\end{corollary}

For an identifier with $p=8$, $u=32$ and $t=32$ ($K=72$ bits), $b=1$, $r=0.5$ and $\alpha=0.45$, Equation~\eqref{eq:nmin} gives $N_{\min}\approx 320$ letters, about $66$ words of our book prose. We call this the design point. At $r=0.5$ half of the raw carrier content is held in reserve, so a text of this length should survive the erasure of about half its carriers. With no reserve, $r\to1/(1+\varepsilon)$, the same identifier needs the decoder to read about $155$ letters of our books and $166$ of our LLM stories (see Section~\ref{sec:evalsel}). 
At $r=0.9$, the carrier sets need $178$ letters; at $r=0.5$, a $40$-bit platform-only identifier needs $178$ letters and a $128$-bit one $569$. These lengths hold for one copy of the payload. In a longer text the rateless code uses the extra carriers as redundancy against edits. The scheme targets paragraph- and article-length text.

\subsubsection{Robustness and keyed placement} \label{sec:robust}
Edits at up to $\beta N$ edit sites add to the loss of the channel. Proposition~\ref{prop:capacity} bounds the per-carrier loss that the rateless code tolerates from both sources together:
\begin{equation}
\kappa \;=\; 1-(1+\varepsilon)\,r ,
\label{eq:pemax}
\end{equation}
about $0.49$ at $r=0.5$. Keeping the placement secret does not render it more tolerant against an adversary who does not know $k$, as the following proposition shows.

\begin{proposition}[Keyed placement does not add robustness]
\label{prop:keying}
Let the carrier code be symbol-symmetric, as a rateless code is, so that the probability of recovery depends only on the number of corrupted carriers. Let $\mathcal{B}_k$ be an adversary that knows $k$ and edits at most $B$ positions. Then there is an adversary $\mathcal{B}_0$ that does not know $k$, edits at most $B$ positions and makes recovery fail with at least the probability achieved by $\mathcal{B}_k$.
\end{proposition}

The key protects the identifier against forgery but it does not protect the mark against removal. Carrier positions follow from the public alphabet, so anyone can find them and fold them to Latin.

\subsubsection{Structural edits: reordering and partial quotation} \label{sec:structural}
Writing $T'=s_1s_2\cdots s_M$ as a concatenation of $M$ sentences, we model selection and reordering as the structural channel
\begin{equation}
\mathcal{S}_{\psi,S}(T')\;=\;s_{\psi(1)}\,s_{\psi(2)}\cdots s_{\psi(|S|)},
\label{eq:structural}
\end{equation}
for a selection $S\subseteq[M]$ and an injection $\psi:[|S|]\to S$ that fixes the order of presentation. 
As we show in the following propositions, \name in unaffected by reordering and deleting sentences only shortens the text, while a cut damages only the carriers next to it.

\begin{proposition}[Invariance to reordering]
\label{prop:perm}
Under the content-anchored placement with span clipping, assume that (i) sentence boundaries are computed on the skeleton $\sigma(\cdot)$; (ii) the clipped span excludes the whitespace that separates one sentence from the next; and (iii) the segmentation of $\mathcal{S}_{\psi,S}(T')$ recovers the same sentences for every $\psi$. Then the multiset of (index, bit) pairs that the decoder collects from the interior of the retained text is a function of the selection $S$ alone. If the decoder reads every retained carrier, $D_k(\mathcal{S}_{\psi,S}(T'))=D_k(\mathcal{S}_{\mathrm{id},S}(T'))$. A decoder that skips the carriers within $W_{\mathrm{ctx}}/2$ of the ends of the received text admits an order dependence confined to $O(\alpha_c W_{\mathrm{ctx}})$ carriers at the two ends, a constant boundary layer that does not grow with the number of sentences.
\end{proposition}

\begin{proof}[Proof of Proposition~\ref{prop:perm}]
Segmentation is computed on $\sigma(\tilde T)$. By (i) and (iii) the decoder recovers the sentences that the embedder used, in whatever order they arrive. The decoder's input is the multiset $\{(\mathrm{idx}(c),\mathrm{bit}(c)):c\ \text{a retained carrier}\}$. For a carrier $c$ in sentence $s_j$, $\mathrm{idx}(c)=H_k(\mathrm{ctx}(c))\bmod\Lambda$, with $\mathrm{ctx}(c)$ the normalized window clipped at the boundaries of $s_j$. By (ii) that window is a function of $s_j$ alone. Without (ii) it is not: a segmenter that conforms to UAX\,\#29~\cite{uax29} assigns the space following a sentence to that sentence, so $s_j$ acquires a trailing blank exactly when another sentence follows it, and its last $W_{\mathrm{ctx}}/2$ carriers would read one context inside a document and another at its end. Given (ii), $\mathrm{ctx}(c)$ is unchanged by the deletion of any other sentence and by any reordering, neither of which alters the interior of $s_j$. The bit $\mathrm{bit}(c)$ is the code point at $c$, likewise untouched. The multiset is thus a function of $S$ alone. A fountain decoder is a function of an unordered collection of coded bits, and $D_k$ depends only on that multiset. For the last statement, which carriers lie within $W_{\mathrm{ctx}}/2$ of the ends of the received text is determined by the sentences presented first and last. A decoder that skips them reads a multiset that depends on $\psi$ through those $O(\alpha_c W_{\mathrm{ctx}})$ positions and through nothing else.
\end{proof}

Span clipping is the reason the number of moved sentences does not matter. Without clipping, a window that crosses a boundary is corrupted whenever the neighboring sentence moves, and reordering $M$ sentences would desynchronize the $O(\alpha_c W_{\mathrm{ctx}}M)$ carriers next to boundaries. This damage would be survivable, but it would grow with the document. With clipping only $O(\alpha_c W_{\mathrm{ctx}})$ carriers are affected, independently of $M$.

\begin{proposition}[Selection threshold]
\label{prop:excerpt}
Let a sentence-aligned structural edit retain sentences totaling $N_S$ letters, hence $L_S\approx\alpha N_S$ carriers with $\nu(L_S)$ distinct indices. Recovery of a $K$-bit payload succeeds with high probability once
\begin{equation}
b\,\nu(L_S)\;\ge\;(1+\varepsilon)K ,
\label{eq:excerpt}
\end{equation}
and is impossible when $b\,\nu(L_S)<K$.
\end{proposition}

Contexts repeat in natural language and indices collide modulo $\Lambda$, so $\nu(\ell)\le\ell$. Near the threshold $\nu(\ell)/\ell=0.96$, and we absorb this ratio into the code rate $r$. In the regime $\nu(\ell)=\ell$, condition~\eqref{eq:excerpt} reduces to $N_S\ge N_{\min}$: any excerpt at least as long as the minimum markable length can be attributed, regardless of how it was assembled. For the $72$-bit identifier at the design point this means $320$ letters of retained text, taken as one paragraph or as sentences from anywhere in the document.

\begin{proof}[Proof of Proposition~\ref{prop:excerpt}]
By Proposition~\ref{prop:perm} the decoder sees exactly the (index, bit) pairs of the retained carriers, and only the retained set matters. A rateless code recovers $K$ payload bits from any $(1+\varepsilon)K$ coded bits on distinct indices. Repeated indices carry the same coded bit, are combined by majority vote and count once (Section~\ref{sec:placement}). The retained carriers therefore provide $\nu(L_S)$ usable coded symbols of $b$ bits each, which gives Equation~\eqref{eq:excerpt}. The impossibility part is Proposition~\ref{prop:capacity} applied to the excerpt: its $\nu(L_S)$ distinct indices hold $b\,\nu(L_S)$ bits in total, and no extractor outputs more information about $m$ than its input contains. The index is a keyed hash of local content; in the random-oracle model the indices of the retained carriers form a uniform sample of the index range, whose distribution depends only on how many carriers are retained. Deleting one run of $d$ carriers and deleting $d$ scattered carriers lead to the same distribution of the collected index set, so losses that come in bursts in text positions are i.i.d.\ in index positions. The achievability part of Proposition~\ref{prop:capacity} holds for arbitrarily correlated losses of this kind.
\end{proof}

\begin{proposition}[Fragment locality]
\label{prop:fragment}
Let $\tilde T$ be any concatenation, in any order, of $F$ pairwise-disjoint contiguous fragments of $T'$ totaling $L_S$ carriers.
(i) If every cut falls at a sentence boundary of $T'$, no carrier is damaged and Propositions~\ref{prop:perm} and~\ref{prop:excerpt} apply unchanged: $F$ is unbounded.
(ii) If cuts fall inside sentences, only carriers whose context window contains a cut are damaged, at most $W_{\mathrm{ctx}}\alpha_c$ per cut and $2W_{\mathrm{ctx}}\alpha_c$ per fragment. With an outer code of error-correction radius $\theta$, recovery holds while
\begin{equation}
F\;\le\;\frac{\theta\,L_S}{2\,W_{\mathrm{ctx}}\,\alpha_c}.
\label{eq:fragbound}
\end{equation}
\end{proposition}

Part~(i) covers quotations cut at sentence boundaries. A post quoted as sentences $1,2,5,7,14$, reordered and split across three replies, reaches the decoder as one excerpt of the same total length. Part~(ii) covers a sentence quoted in part, where a cut damages only the carriers in its own window. On long texts, the bound~\eqref{eq:fragbound} is pessimistic, because a wrong vote on an index where there are many correct votes is outvoted. Near the threshold, where there are no spare votes, a decoder that searches for a consistent subset of equations takes the place of the outer code. Table~\ref{tab:structural} in Appendix~\ref{app:extra} summarizes the cost of each structural edit.

\subsubsection{Capacity--robustness frontier} \label{sec:frontier}
Eliminating $r$ between $K=b\,r\,\alpha N$, the Equation~\eqref{eq:pemax}, for a text of $N$ letters, yields:
\begin{equation}
K\;\approx\;\frac{b\,\alpha N}{1+\varepsilon}\,\big(1-\kappa\big).
\label{eq:frontier}
\end{equation}
Payload and tolerated loss trade off linearly: each additional $0.1$ of tolerated loss costs $0.1\,b\alpha N/(1+\varepsilon)$ bits of payload.

\subsubsection{Detectability and the achievable region} \label{sec:detect}
\begin{definition}[Covert security]
\label{def:covert}
Against a detector class $\mathcal{D}$, the scheme is $\epsilon$-covert if $\big|\Pr[D(E_k(T,m))=1]-\Pr[D(T)=1]\big|\le\epsilon$ for all $D\in\mathcal{D}$~\cite{cachin,hla}.
\end{definition}
Let $q_i$ be the probability that carrier $i$ appears in its non-Latin variant in cover text. A marked word mixes scripts within the word, whereas multilingual text mixes them between words~\cite{tr39}.

\begin{proof}[Proof of Proposition~\ref{prop:fragment}]
(i) By span clipping, a carrier's window lies inside its own sentence. If that sentence is retained whole, the window content, hence the index, is identical to the original, and the code point is untouched.
(ii) Fix a cut. A carrier whose window does not contain the cut point reads the same characters as in $T'$, because clipping already confined the window to the sentence and the cut lies outside the window; its index is correct even when the rest of its sentence is gone. Only carriers within one window width of the cut read a different context. There are at most $W_{\mathrm{ctx}}$ such character positions, of which a fraction $\alpha_c$ are carriers. Each such carrier yields a wrong index, which either lands on an unused index or collides with a used one. Bounding both cases by the count and requiring the total to lie inside the radius of the outer code gives Equation~\eqref{eq:fragbound}.
\end{proof}

\begin{proposition}[Detectability converse]
\label{prop:detect}
Let an embedding set carrier $i$ to its non-Latin variant with probability $\delta_i$, independently across carriers. The optimal test distinguishes marked from cover text with advantage equal to the total variation distance of the two variant distributions, which is bounded below by
\begin{equation}
\mathrm{Adv}\;\ge\;1-\prod_i\Big(\sqrt{\delta_i q_i}+\sqrt{(1-\delta_i)(1-q_i)}\Big),
\label{eq:detect}
\end{equation}
and tends to $1$ as the number of carriers with $\delta_i\neq q_i$ grows. An $\epsilon$-covert scheme therefore has covert capacity at most $\sum_i H(q_i)+O(L\epsilon)$, the code-point entropy of the cover. For cross-script confusables in monolingual text $q_i\approx0$ and $\sum_i H(q_i)\approx0$. Therefore, the bound carries an $O(L\epsilon)$ slack and is informative for $\epsilon=o(1/L)$.
\end{proposition}

\begin{proof}[Proof of Proposition~\ref{prop:detect}]
The maximum distinguishing advantage of any test between two distributions equals their total variation distance. For the product distributions of the carrier variants, $\mathrm{TV}\ge 1-\mathrm{BC}$ with Bhattacharyya coefficient $\mathrm{BC}=\prod_i\big(\sqrt{\delta_i q_i}+\sqrt{(1-\delta_i)(1-q_i)}\big)$, which gives Equation~\eqref{eq:detect}. Each factor is strictly below $1$ when $\delta_i\neq q_i$, hence $\mathrm{BC}\to0$ and $\mathrm{TV}\to1$. Covertness ($\mathrm{TV}\le\epsilon$) therefore forces every $\delta_i$ within $O(\epsilon)$ of $q_i$, and the entropy conveyed per carrier cannot exceed $H(q_i)+O(\epsilon)$ by the information-theoretic steganography bound~\cite{cachin,hla}; with $q_i\approx0$ each $H(q_i)\approx0$.
\end{proof}

\begin{proposition}[Achievable region and covert limit]
\label{prop:region}
For the weight-$w$ binary carrier ensemble over $L$ carriers on an erasure channel of rate $p_e$:
(i) the recoverable payload obeys $K\le(1-p_e)\,L\,H_2(w)$, attained within $(1+\varepsilon)$ by a rateless code over the ensemble; (ii) in monolingual cover ($q_i\approx0$) the mixed-script test distinguishes the emitted text with advantage at least $1-(1-w)^{L/2}$; (iii) holding that advantage below a constant $a<1$ forces $w=O(1/L)$ and a covert payload of $O(\log L)$ bits.
\end{proposition}

At $w=\tfrac12$ the region recovers the full binary payload of Proposition~\ref{prop:capacity}; as $w\to0$ the payload vanishes. A covert mark carries $O(\log L)$ bits, and a fixed identifier of $p+u+t$ bits would need covert text whose length grows exponentially in the identifier size. For this reason, \name does not try to be covert against a mixed-script test and uses the maximum-weight code, $w=\tfrac12$. 
\begin{proof}[Proof of Proposition~\ref{prop:region}]
(i) Each carrier is non-Latin independently with probability $w$. Its state carries at most $H_2(w)$ bits and the emitted carrier vector at most $L H_2(w)$. Erasures leave $(1-p_e)L$ independent carriers, and by the counting bound of Proposition~\ref{prop:capacity} the recoverable payload is at most their entropy $(1-p_e)L H_2(w)$, met within $(1+\varepsilon)$ by a rateless code over the ensemble.
(ii) Setting $\delta_i=w$ and $q_i\approx0$ in Equation~\eqref{eq:detect} gives the Bhattacharyya coefficient $\prod_i\sqrt{1-w}=(1-w)^{L/2}$, whence $\mathrm{Adv}\ge 1-(1-w)^{L/2}$, under the same independent-carrier model as in part~(i).
(iii) $\mathrm{Adv}\le a$ requires $(1-w)^{L/2}\ge1-a$, i.e., $w\le\frac{2}{L}\ln\frac{1}{1-a}\,(1+o(1))$; then $H_2(w)=O\!\big(\tfrac{\log L}{L}\big)$ and $K\le(1-p_e)L H_2(w)=O(\log L)$.
\end{proof}

\subsubsection{Security} \label{sec:security}
\begin{proposition}[Unforgeability]
\label{prop:forge}
Let the authenticated encryption have ciphertext-integrity advantage $\mathrm{Adv}^{\mathrm{AEAD}}$. An adversary without $k$ that makes $q$ verification attempts forges a valid identifier with probability at most $q\,2^{-t}+\mathrm{Adv}^{\mathrm{AEAD}}$.
\end{proposition}

For the truncated-tag construction, $\mathrm{Adv}^{\mathrm{AEAD}}$ is the PRF advantage of the primitive from which the tag and the keystream are derived. Every tag check counts as a query, including each epoch key tried on a candidate and each candidate that a searching decoder submits. For $t=32$ and an offline or rate-limited verifier, the forgery probability is $2^{-32}\approx 2.3\times10^{-10}$ per tag check. An open verifier should use $t=64$, which adds $32$ payload bits, about $140$ letters at $r=0.5$ by Corollary~\ref{cor:nmin}. Acceptance requires a verifying tag, which an adversary without $k$ cannot produce for a new payload. This rules out spoofing, in which an adversary attaches a valid-looking attribution to text it did not create~\cite{wmstealing}. On a spliced sample the extractor returns the identifier whose carriers dominate the sample, or $\bot$; decoding groups of sentences separately attributes each span to its own identifier.

\begin{proof}[Proof of Proposition~\ref{prop:forge}]
A forger that produces a valid tag on a fresh (ciphertext, identifier) pair breaks the ciphertext integrity of the authenticated encryption, which contributes $\mathrm{Adv}^{\mathrm{AEAD}}$. Otherwise each of the $q$ attempts must guess the $t$-bit tag and is accepted with probability $2^{-t}$, and a union bound gives $q\,2^{-t}$. For the truncated-tag verification rule of Section~\ref{sec:payload}, replace $\mathrm{MAC}_k$ and $F_k$ by independent random functions, at the cost of their combined PRF advantage, which is the $\mathrm{Adv}^{\mathrm{AEAD}}$ term in this mode. For every candidate $(\mathrm{ct}',\tau')$ the recomputed truncated tag is then a fresh uniform $t$-bit string, the acceptance probability is exactly $2^{-t}$, and the union bound gives $q\,2^{-t}$. When the verifier tries $\mathcal{E}$ epoch keys per candidate, $q$ counts every (candidate, epoch) pair.
\end{proof}

\begin{proposition}[Public verifiability and private attribution]
\label{prop:pubver}
Under EUF-CMA security of the signature~\cite{gmr}, collision resistance of $H$, and IND-CPA security of the encryption: (i) any verifier holding $\mathrm{pk}$ accepts every honestly marked text; (ii) no adversary without $\mathrm{sk}$ produces text of different content that verifies as the platform's, except with negligible probability; (iii) the mark is non-transferable, as a valid signature fails on content whose normalized hash differs from $h$; and (iv) the user identifier is hidden from every party without $K_P$.
\end{proposition}

\begin{proof}[Proof of Proposition~\ref{prop:pubver}]
(i) is the completeness of the signature scheme. For (ii) and (iii), text of content hash $h'\neq h$ that verifies yields either a signature forgery on $\mathrm{pid}\,\Vert\,h'$, contradicting EUF-CMA security, or a collision $H(\sigma(T''))=h$, contradicting collision resistance. (iv) is the IND-CPA security of $\mathrm{Enc}_{K_P}$.
\end{proof}

Symmetric text watermarks need the issuers secret to attribute a text or a trusted third party. In the asymmetric mode an investigator confirms the platform of origin with the public key alone, and the user identity stays under the platform's control.

\begin{table}[t]
\caption{Statistics of the two corpora.}
\label{tab:corpora}
\centering
\footnotesize
\setlength{\tabcolsep}{4pt}
\begin{tabular}{lrr}
\toprule
 & Books & LLM stories \\
\midrule
Words in the corpus & 475{,}178 & 56{,}658 \\
Sentences kept / Total sentences & 73.76 & 49.98 \\ 
Letters per kept sentence & 142.6 & 91.3 \\
Carriers per kept sentence & 70.7 & 42.1 \\
Documents near the threshold & 40 & 40 \\
Letters per document & 4004--4267 & 3916--4146 \\
Sentences per document (mean) & 30.5 & 44.5 \\
Carrier fraction of letters $\alpha$ (pooled) & 0.4999 & 0.4681 \\
$\alpha$ on the embedder's carrier set & 0.4956 & 0.4609 \\
Carriers per character $\alpha_c$ & 0.389 & 0.354 \\
Carriers skipped at the two ends & 6.20 & 5.58 \\
$\nu/L$, shortest document ($L$) & 0.943 (105) & 0.973 (74) \\
$\nu/L$, 4000-letter document ($L$) & 0.446 (1999) & 0.460 (1834) \\
\bottomrule
\end{tabular}

\end{table}
\section{Experimental Setup} \label{sec:setup}\label{sec:evalsetup}
This section describes the experimental setup used to evaluate \name.
\subsection{Datasets} \label{sec:data}
We evaluate \name on two English corpora details of which are shown in Table~\ref{tab:corpora}. We chose three public-domain books from Project Gutenberg~\cite{gutenberg}: Pride and Prejudice (Austen), On the Origin of Species (Darwin) and The Federalist Papers. Moreover we generated $40$ short stories using Claude Haiku~5.5. We asked for eight genres (mystery, science fiction, fantasy, historical, literary, adventure, fable and humor) with five premises each, using the prompt ``Write a short story of about 1,400 words. Premise: \textless premise\textgreater''. The LLM generated stories are meant to represent the text an LLM platform would watermark. 

Experiments that select sentences draw them from spans of $40$ to $400$ letters that end in punctuation, are at least three-quarters lower-case and survive a segmentation round trip. Applied to the full text files, this filter keeps 73.76$\%$ of the spans of the books and 50$\%$ of the stories, most of the removed story spans are shorter than $40$ letters. For the experiments near the threshold we cut $40$ documents (a run of consecutive kept sentences) of about $4000$ letters from each corpus. The book documents are cut from the body of each book at evenly spaced offsets, and each is marked under one of $8$ keys. The story documents are the kept sentences at the start of each story, story $d$ marked under key $d \bmod 8$. 

The carrier fraction of letters is about $0.5$ in the books and $0.47$ in the stories, against the $0.49$ predicted from letter frequencies (see Section~\ref{sec:capacity}). 
On the carrier set that the embedder writes, $\alpha=0.49$ for books and $0.46$ for the LLM generated stories. Both pooled values lie above the design value $\alpha=0.45$.
In the books, distinct context windows stay between $0.994$ and $1.000$ of $L$ for documents of $218$ to $4104$ letters, while distinct indices fall from $0.94$ to $0.45$ of $L$. This is the occupancy of $\Lambda$ bins under $L$ throws: the estimate $\Lambda\big(1-(1-1/\Lambda)^{L}\big)$ is within $4\%$ of the measured counts ($3\%$ on the stories, whose windows stay between $0.98$ and $1$ of $L$). Near the threshold ($L\approx76$) the occupancy model gives $\nu/L=0.96$. The single-document-fits place the $50\%$ point at $73.0$ distinct indices for $75.9$ read carriers in the books and at $73.1$ for $76.3$ in the stories, so about $0.96$ for both. Given that, we size the scheme with $\nu/L=0.96$.

\subsection{Evaluation metrics} \label{sec:protocol}
The recovery rate is the fraction of trials in which the extractor returns the correct $(\mathrm{pid},\mathrm{uid})$. The false-accept rate is the fraction of unmarked or foreign samples for which the extractor returns any identifier. The $50\%$ point is the excerpt length at which a logistic fit of recovery against received letters crosses $0.5$ and we report it with a bootstrap $95\%$ interval ($1000$ resamples). As a measure of cost we count tag checks, i.e., the verification queries that the extractor spends per sample.
\section{Evaluation}\label{sec:eval}
This section evaluates \name along three axes: 1) copy-paste transport; 2) structural edits; 3) character-level edits and erasures.

\subsection{Copy-paste channel} \label{sec:evalchannel}
We pasted watermarked text into the following $18$ platforms: X/Twitter, Facebook, Facebook Messenger, Instagram (comments and messages), WhatsApp Web, Signal, Telegram, Discord, Microsoft Teams, Gmail, Proton Mail, Tuta Mail, Google Docs, Google Sheets, Apple Notes, NeoVim and KDE Connect clipboard sharing. Subsequently, we ran the extractor, and on each run, none of the $18$ platforms folded or rewrote the carriers.

\subsection{Recovery threshold} \label{sec:evalsel}
As shown in Section~\ref{sec:evalsetup} the code requires $(1+\varepsilon)K=73.6$ distinct coded bits. For $\nu/L=0.96$ it means $76.7$ carriers, that is, $155$ letters read by the decoder at $\alpha=0.49$ in the books and $166$ letters at $\alpha=0.46$ in the stories. 
We compared the prediction with a logistic fit of the measured recovery and the results are shown in Table~\ref{tab:threshold}.

\begin{table}[t]
\caption{Predicted and measured $50\%$ recovery points (bootstrap $95\%$ intervals).}
\label{tab:threshold}
\centering
\scriptsize
\setlength{\tabcolsep}{3pt}
\begin{tabular}{lcc}
\toprule
 & Books & LLM stories \\
\midrule
Predicted, letters read & $155$ & $166$ \\
Predicted, with boundary layer & $167$ & $178$ \\
\midrule
Character excerpts & $167.3$ [$166.2$, $168.5$] & $177.6$ [$176.4$, $178.9$] \\
Sentence selections & $168.4$ [$166.4$, $170.2$] & $177.8$ [$176.1$, $179.3$] \\
Outer code (b), excerpts & $275.4$ [$274.1$, $276.7$] & $292.7$ [$291.3$, $294.2$] \\
One doc., read carriers & $75.9$ [$75.5$, $76.2$] & $76.3$ [$75.9$, $76.7$] \\
One doc., distinct indices & $73.0$ [$72.7$, $73.3$] & $73.1$ [$72.8$, $73.4$] \\
\bottomrule
\end{tabular}

\end{table}
On one document of each corpus, the $50\%$ point lies at $73.0$ and $73.1$ distinct read indices and at $75.9$ and $76.3$ read carriers, against the predicted $73.6$ and $76.7$. Over the $40$ documents, the $50\%$ point lies at $167$ received letters for character excerpts, at $168$ for sentence selections in the books, and at $178$ for both selection types in the stories. The difference between the letters that the decoder reads and the letters it receives is the boundary layer defined in Section~\ref{sec:ecc}. The decoder skips $6.2$ carriers at the two ends of a book excerpt and $5.6$ at the ends of a story excerpt, about $12$ letters in both cases, adding them brings the predictions to the measured $167$ and $178$ letters.
The transition between the two regimes is sharp (see Table~\ref{tab:regime}). In the books, recovery rises from $0.11$ at $155$ letters to $0.99$ at $195$, and the stories follow the same curve shifted by about ten letters. In Table~\ref{tab:structure} we analyze the impact of counting the threshold in sentences. On average a kept sentence has $142.6$ letters in the books and $91.3$ in the stories (Table~\ref{tab:corpora}). Two kept sentences recover $0.84$ (scattered) and $0.87$ (contiguous) in the books, but only $0.52$ and $0.50$ in the stories, while three sentences recover at least $0.92$ in both corpora. The LLM text is harder to mark because it has fewer carriers per letter ($\alpha=0.4609$ against $0.4956$) and is written in shorter sentences, so that a quotation of a given number of sentences holds fewer letters; the two effects compound. The decoder itself needs the same number of distinct indices on both corpora.
\begin{table}[t]
\caption{Reordering at the threshold ($300$ documents per length; $1000$ permutations of $3$--$5$ short sentences, $230$ for five book sentences), and recovery from two or three kept sentences ($40$ documents, $1280$ trials per cell).}
\label{tab:structure}
\centering
\footnotesize
\setlength{\tabcolsep}{3pt}
\begin{tabular}{lcc}
\toprule
 & Books & LLM stories \\
\midrule
Order changes verdict, 160--300 letters & $0/1524$ & $0/1560$ \\
Order changes verdict, 3 short sentences & $8/1000$ & $1/1000$ \\
Order changes verdict, 4 short sentences & $6/1000$ & $0/1000$ \\
Order changes verdict, 5 short sentences & $1/230$ & $0/1000$ \\
\midrule
2 sentences kept, scattered & $0.84$ & $0.52$ \\
2 sentences kept, contiguous & $0.87$ & $0.50$ \\
3 sentences kept, scattered & $0.99$ & $0.93$ \\
3 sentences kept, contiguous & $0.98$ & $0.92$ \\
\bottomrule
\end{tabular}

\end{table}
\subsection{Structural edits} \label{sec:evalperm}
We evaluate the robustness of \name watermarking technique to structural edits such as reordering, partial quotation and character edits.  
\subsubsection{Reordering.} We permuted at random sentences of around 4000-letters, all $256$ permutations of each were recovered. As long documents have greater redundancy, to test the limit we marked $300$ documents at each of six lengths from $160$ to $300$ letters. The embedder accepted $227$ out of $300$ book documents at $160$ letters and accepted all from $205$ onwards. As shown in Table~\ref{tab:structure}, permutations did not change the verdict in any permutable book or story documents.
Order dependence is present only when the decoder skips the carriers within $W_{\mathrm{ctx}}/2$ of the ends of the received text (Proposition~\ref{prop:perm}). Documents of three to five short sentences totaling $160$ to $230$ letters changed verdict in $8$ of $1000$, $6$ of $1000$ and $1$ of $230$ book permutations, and in $1$ of $1000$ three-sentence story permutations. In such cases the carriers skipped at the two ends are a significative fraction of the total, and a permutation changes which carriers they are.

\subsubsection{Partial quotation.} Scattered and contiguous selections of the same number of sentences behave similarly. This is consistent with Proposition~\ref{prop:excerpt}, according to which recovery depends only on the retained length and not on where in the document the retained text comes from.
\begin{table}[t]
\caption{Recovery of the $40$ documents after scattered edits, by decoder ($320$ trials per cell).}
\label{tab:edits}
\centering
\footnotesize
\setlength{\tabcolsep}{3.5pt}
\begin{tabular}{llccccc}
\toprule
 & Edits & $100$ & $150$ & $200$ & $250$ & $300$ \\
\midrule
\multicolumn{7}{l}{{Books}} \\
Elim. & subst. & $1.00$ & $0.95$ & $0.52$ & $0.13$ & $0.01$ \\
Elim. & deleted & $1.00$ & $0.93$ & $0.53$ & $0.19$ & $0.02$ \\
Elim. & inserted & $1.00$ & $0.98$ & $0.64$ & $0.16$ & $0.03$ \\
(a) & subst. & $1.00$ & $1.00$ & $0.99$ & $0.65$ & $0.08$ \\
(a) & deleted & $1.00$ & $1.00$ & $1.00$ & $0.68$ & $0.07$ \\
(a) & inserted & $1.00$ & $1.00$ & $1.00$ & $0.70$ & $0.12$ \\
\midrule
\multicolumn{7}{l}{{LLM stories}} \\
Elim. & subst. & $1.00$ & $0.92$ & $0.45$ & $0.09$ & $0.02$ \\
Elim. & deleted & $0.99$ & $0.91$ & $0.49$ & $0.09$ & $0.01$ \\
Elim. & inserted & $1.00$ & $0.93$ & $0.47$ & $0.11$ & $0.01$ \\
(a) & subst. & $1.00$ & $1.00$ & $1.00$ & $0.63$ & $0.08$ \\
(a) & deleted & $1.00$ & $1.00$ & $0.99$ & $0.66$ & $0.07$ \\
(a) & inserted & $1.00$ & $1.00$ & $1.00$ & $0.71$ & $0.08$ \\
\bottomrule
\end{tabular}

\end{table}
\subsubsection{Fragmentation.} We kept half of each single document, split it into $F$ pieces, either at sentence boundaries or into equal pieces at character offsets, shuffled the pieces and measured recovery. Cuts at sentence boundaries had no effect (part~(i) of Proposition~\ref{prop:fragment}). Recovery remains at $1.00$ for every $F$ up to $16$ on the half documents. For the whole documents, one piece per sentence ($F= 25$ for the book documents and $F=46$ for the LLM stories) was enough to allow for $100\%$ recovery. Cuts inside sentences did not hurt at this length. With $16$ pieces of the half documents, and $64$ of the whole documents, recovery stayed at $1.00$. 
Near the threshold, however, there are not enough spare votes to outvote the carriers next to a cut. The results are presented in Table~\ref{tab:regime}. At $230$ letters, $F=2$ leaves recovery at $0.46$ and $F=4$ at $0.03$; at the $320$-letter design point, $F=2,4,8$ leave $0.73$, $0.12$ and $0$. By Proposition~\ref{prop:fragment}, each internal cut creates a join where about $W_{\mathrm{ctx}}\alpha_c\approx6.6$ carriers read a context that does not occur in the original. These carriers are mis-indexed rather than erased, and Section~\ref{sec:evaldecode} shows that a decoder that tolerates errors removes most of this cost.

\subsubsection{Effect of surrounding whitespace.} The boundary layer is measured from the two ends of the sample. When an excerpt arrives with surrounding whitespace, the skipped region covers the whitespace instead of letters, and the carriers whose windows were cut by the excerpt are read and mis-indexed. Table~\ref{tab:whitespace} in Appendix~\ref{app:extra} displays this effect on $240$-character excerpts. A newline before or after the excerpt lowers recovery from $0.97$ to $0.89$ in the books, and two spaces at both ends lower it to $0.81$; the stories, which start from $0.81$ at this length, lose a similar amount. Trimming the sample before decoding restores the bare-excerpt rate in every case, to within two of $640$ trials. 

\subsection{Character-level edits} \label{sec:evaledit}
Recasing has no effect at any budget, the context is lower-cased before hashing and a letter is a carrier only if both of its cases are. Scattered substitutions, deletions and insertions behave alike. Table~\ref{tab:edits} presents the recovery rate after each attack, on 2 decoders for $100-300$ edited characters. With the elimination decoder, recovery on the $40$ book documents falls from $0.93$--$0.98$ at $150$ edits to $0.52$--$0.64$ at $200$ and $0.13$--$0.19$ at $250$, i.e., it collapses near $200$ scattered edits, about $5\%$ of the letters. The LLM stories collapse near the same budget, slightly earlier, namely $0.45$--$0.49$ at $200$ edits. Although insertions and deletions shift every later character, they do no more harm than substitutions, since the placement depends only on local content.

\begin{figure}[t]
\centering
\includegraphics[width=0.8\columnwidth]{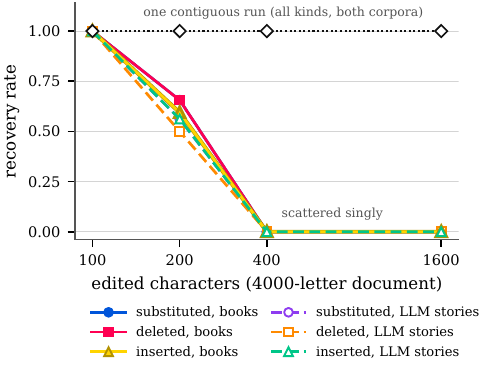}
\caption{Recovery of the single documents after the same number of edited characters ($32$ trials per point). Coloured series report scattered edits, solid for the books and dashed for the LLM stories; the black line reports one contiguous run.}
\label{fig:editsites}
\end{figure}

\subsubsection{Edit sites.} 
\begin{figure}[t]
\centering
\includegraphics[width=0.8\columnwidth]{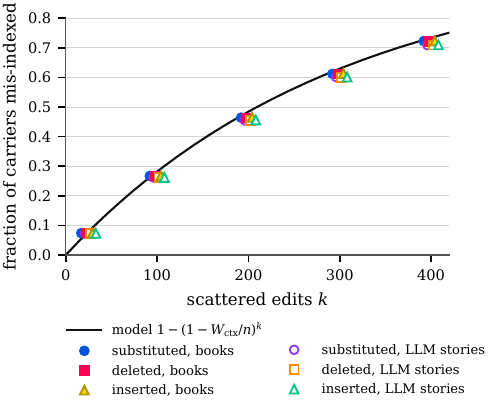}
\caption{Fraction of carriers mis-indexed by $k$ scattered edits. The black line reports the window-overlap model, whereas markers report measurements on the books and LLM stories; markers shifted horizontally for visibility.}
\label{fig:misindex}
\end{figure}

Figure~\ref{fig:editsites} shows that the same number of edited characters does far less damage in one run. On the single documents, $400$ characters edited individually prevent recovery in both corpora, while one run over $4000$ characters leaves recovery at $1.00$. An edit damages the carriers whose window it overlaps (about $W_{\mathrm{ctx}}\alpha_c\approx6.6$ of them), and a run of any length overlaps one window at each of its two ends. With $k$ scattered edits in a document of $n$ characters, the expected fraction of carriers with an edit in their window is $1-(1-W_{\mathrm{ctx}}/n)^{k}$. Figure~\ref{fig:misindex} compares this model with the measured fraction of mis-indexed carriers. For both the story and book documents of around $5150$ characters, the model gives an estimation within an $0.02$ margin of error
A run still removes text, and once little text is left on either side of it, recovery is limited by the threshold. Leaving $800$, $350$, $300$ and $200$ characters of the single book document untouched by a run gives recovery of $0.96$, $0.55$, $0.43$ and $0$ ($0.95$, $0.55$, $0.33$ and $0$ for the story). The $50\%$ point lies between $330$ and $340$ untouched characters, about $265$ letters, above the bare threshold. Our explanation is that the two ends of the run add mis-indexed carriers on top of the lost text.

\subsection{Erasures and errors} \label{sec:evalerasure}
\begin{figure}[t]
\centering
\includegraphics[width=0.8\columnwidth]{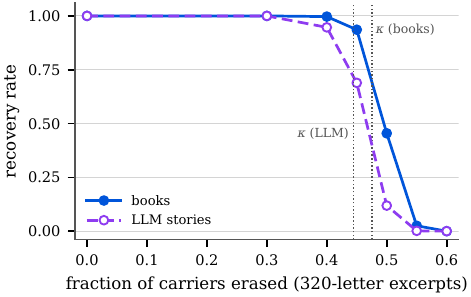}
\caption{Recovery of $320$-letter excerpts when a growing fraction of their carriers is erased ($640$ trials per point), for the books and the LLM stories. Dotted vertical lines mark the predicted tolerance $\kappa$ of each corpus.}
\label{fig:erasure}
\end{figure}

Pure erasures drop votes and leave every context intact, which is the regime of Proposition~\ref{prop:capacity}. Table~\ref{tab:erasure} (Appendix~\ref{app:extra}) and Figure~\ref{fig:erasure}, present erasures of a growing fraction of the carriers of $320$-letter excerpts. In the books, recovery stays at $1.00$ up to an erasure rate of $0.40$, and drops to $0.94$ at $0.45$ and to $0.45$ at $0.50$. The $50\%$ point lies at $73$ surviving distinct indices, close to the predicted value. Its predicted tolerance $\kappa=1-73.6/140.4=0.48$ falls on the measured transition. A $320$-letter story excerpt holds $132$ distinct indices instead of $140$, because of its lower carrier fraction. Its predicted tolerance is therefore lower, $\kappa=0.44$, and its recovery falls earlier ($0.69$ at $0.45$ and $0.12$ at $0.50$), again with the $50\%$ point near $73$ surviving indices. On the single documents ($256$ trials per cell), $640$-letter excerpts stay at $1.00$ up to an erasure rate of $0.70$ in both corpora.
With the elimination decoder, one, two and four scattered substitutions at the same length leave recovery at $0.69$, $0.32$ and $0.04$.
Four substitutions mis-index about $26$ carriers, about a third of the $76$ carriers that the code tolerates as erasures at this length, and yet they almost always prevent recovery. Comparing the $50\%$ points, about $1.5$ scattered substitutions, i.e., about ten mis-indexed carriers, cost as much as the erasure of about $80$ carriers, half of the excerpt. A mis-indexed carrier contributes a wrong equation, and elimination has no way to tell it from a correct one. Near the threshold, an error thus costs close to an order of magnitude more than an erasure, unless the decoder handles errors explicitly. We investigate this in the next section.
\begin{table*}[t]
\caption{Recovery of $\ell$-letter excerpts near the threshold, by decoder ($40$ documents per corpus, $8$ keys, $640$ trials per cell, the same attacks for every decoder; Wilson $95\%$ half-width $\le0.04$).}
\label{tab:regime}
\centering
\footnotesize
\setlength{\tabcolsep}{3.2pt}
\begin{tabular}{l ccccc c cccc c ccc}
\toprule
& \multicolumn{5}{c}{Elimination decoder} && \multicolumn{4}{c}{(a) Consensus decoder} && \multicolumn{3}{c}{(b) RS outer code} \\
\cmidrule{2-6}\cmidrule{8-11}\cmidrule{13-15}
Excerpt $\ell$ (letters) & $155$ & $195$ & $230$ & $320$ & $530$ && $195$ & $230$ & $320$ & $530$ && $320$ & $400$ & $530$ \\
\midrule
\multicolumn{15}{l}{{Books}} \\
Untouched & $0.11$ & $0.99$ & $1.00$ & $1.00$ & $1.00$ && $0.99$ & $1.00$ & $1.00$ & $1.00$ && $1.00$ & $1.00$ & $1.00$ \\
Cut into $F=2$ & $0.03$ & $0.29$ & $0.46$ & $0.73$ & $0.96$ && $0.66$ & $0.99$ & $1.00$ & $1.00$ && $0.28$ & $0.54$ & $0.77$ \\
Cut into $F=4$ & $0.00$ & $0.01$ & $0.03$ & $0.12$ & $0.52$ && $0.06$ & $0.34$ & $0.92$ & $0.98$ && $0.01$ & $0.05$ & $0.17$ \\
Cut into $F=8$ & $0.00$ & $0.00$ & $0.00$ & $0.00$ & $0.05$ && $0.00$ & $0.00$ & $0.09$ & $0.83$ && $0.00$ & $0.00$ & $0.00$ \\
$1$ scattered substitution & $0.00$ & $0.14$ & $0.39$ & $0.69$ & $0.96$ && $0.62$ & $0.97$ & $1.00$ & $1.00$ && $0.21$ & $0.43$ & $0.71$ \\
$2$ scattered substitutions & $0.00$ & $0.01$ & $0.08$ & $0.32$ & $0.73$ && $0.17$ & $0.69$ & $0.98$ & $1.00$ && $0.04$ & $0.13$ & $0.31$ \\
$4$ scattered substitutions & $0.00$ & $0.00$ & $0.00$ & $0.04$ & $0.30$ && $0.02$ & $0.13$ & $0.71$ & $0.94$ && $0.00$ & $0.01$ & $0.05$ \\
One run of $10$ characters & $0.00$ & $0.08$ & $0.34$ & $0.69$ & $0.93$ && $0.48$ & $0.95$ & $0.99$ & $1.00$ && $0.17$ & $0.40$ & $0.70$ \\
One run of $50$ characters & $0.00$ & $0.01$ & $0.09$ & $0.56$ & $0.93$ && $0.02$ & $0.53$ & $1.00$ & $1.00$ && $0.04$ & $0.29$ & $0.63$ \\
\midrule
\multicolumn{15}{l}{{LLM stories}} \\
Untouched & $0.03$ & $0.91$ & $1.00$ & $1.00$ & $1.00$ && $0.91$ & $1.00$ & $1.00$ & $1.00$ && $0.96$ & $1.00$ & $1.00$ \\
Cut into $F=2$ & $0.01$ & $0.25$ & $0.46$ & $0.72$ & $0.95$ && $0.53$ & $0.94$ & $1.00$ & $1.00$ && $0.27$ & $0.53$ & $0.78$ \\
Cut into $F=4$ & $0.00$ & $0.00$ & $0.03$ & $0.15$ & $0.53$ && $0.03$ & $0.28$ & $0.92$ & $0.98$ && $0.00$ & $0.07$ & $0.18$ \\
Cut into $F=8$ & $0.00$ & $0.00$ & $0.00$ & $0.00$ & $0.03$ && $0.00$ & $0.00$ & $0.07$ & $0.85$ && $0.00$ & $0.00$ & $0.00$ \\
$1$ scattered substitution & $0.00$ & $0.11$ & $0.36$ & $0.69$ & $0.95$ && $0.46$ & $0.94$ & $1.00$ & $1.00$ && $0.14$ & $0.45$ & $0.70$ \\
$2$ scattered substitutions & $0.00$ & $0.02$ & $0.08$ & $0.31$ & $0.73$ && $0.13$ & $0.60$ & $0.97$ & $0.99$ && $0.02$ & $0.13$ & $0.31$ \\
$4$ scattered substitutions & $0.00$ & $0.00$ & $0.00$ & $0.04$ & $0.30$ && $0.01$ & $0.08$ & $0.68$ & $0.95$ && $0.00$ & $0.00$ & $0.03$ \\
One run of $10$ characters & $0.00$ & $0.07$ & $0.28$ & $0.66$ & $0.95$ && $0.30$ & $0.90$ & $1.00$ & $1.00$ && $0.10$ & $0.41$ & $0.70$ \\
One run of $50$ characters & $0.00$ & $0.01$ & $0.06$ & $0.56$ & $0.91$ && $0.01$ & $0.36$ & $1.00$ & $1.00$ && $0.01$ & $0.34$ & $0.65$ \\
\bottomrule
\end{tabular}

\end{table*}
\subsection{Decoder importance} \label{sec:evaldecode}
This section evaluates the elimination decoder and the two error-tolerant variants presented in Section~\ref{sec:decoders} on an equal setting. Table~\ref{tab:regime} reports recovery near the threshold for every decoder and attack, and Figure~\ref{fig:deccut}~(Appendix~\ref{app:extra}) shows its $F=4$ row as a function of the excerpt length.

\subsubsection{Consensus decoding} 
As we can see in Table~\ref{tab:edits}, at the $320$-letter design point, decoder~(a) recovers $1.00$, $0.98$ and $0.71$ of the book excerpts after one, two and four scattered substitutions, where elimination recovers $0.69$, $0.32$ and $0.04$. After cutting the excerpt into $F=2$ and $F=4$ pieces it recovers $1.00$ and $0.92$ (elimination: $0.73$ and $0.12$), and after a $50$-character run $1.00$ (elimination: $0.56$). The stories give nearly the same values, with four substitutions at $0.68$. As expected, the untouched threshold does not move, since decoder~(a) starts with elimination and solves the same equations on clean input; its $50\%$ point lies at $167$ and $178$ letters. On the $40$ documents, decoder~(a) also moves the collapse under scattered edits from about $200$ to about $265$ edits. At $200$ edits, it recovers $0.99$ to $1.00$ of the samples in both corpora, and at $250$ edits, it still recovers $0.63$ to $0.71$
As shown in Table~\ref{tab:interleave}, the same holds for marked text mixed with unmarked text, whose carriers read $0$ and act as errors. With one unmarked sentence among four marked book sentences, decoder~(a) recovers $0.51$ of the samples and elimination $0.18$; a $51$-letter comment prefixed to four marked sentences leaves $0.85$ against $0.46$. The gain is larger on the stories, where elimination almost never recovers the four-sentence samples ($0.03$ and $0.05$) and decoder~(a) recovers $0.42$ and $0.73$.

\begin{table}[t]
\caption{Consecutive marked sentences mixed with unmarked sentences or prefixed by a $51$-letter comment, by decoder ($40$ documents, $320$ trials per cell).}
\label{tab:interleave}
\centering
\footnotesize
\setlength{\tabcolsep}{3.5pt}
\begin{tabular}{lcccc}
\toprule
 & \multicolumn{2}{c}{Books} & \multicolumn{2}{c}{LLM stories} \\
\cmidrule(lr){2-3}\cmidrule(lr){4-5}
Sample & Elim. & (a) & Elim. & (a) \\
\midrule
2 marked sentences alone & $0.76$ & $0.76$ & $0.48$ & $0.48$ \\
4 marked, 1 unmarked & $0.18$ & $0.51$ & $0.03$ & $0.42$ \\
8 marked, 1 unmarked & $0.73$ & $0.89$ & $0.44$ & $0.76$ \\
8 marked, 2 unmarked & $0.48$ & $0.78$ & $0.11$ & $0.42$ \\
2 marked, comment prefixed & $0.05$ & $0.48$ & $0.00$ & $0.11$ \\
4 marked, comment prefixed & $0.46$ & $0.85$ & $0.05$ & $0.73$ \\
\bottomrule
\end{tabular}

\end{table}

This improvement comes at the cost of additional verification queries. Decoder~(a) uses $4.70$ tag checks per marked book sample with one substitution (at most $9$), and $18.07$ per unmarked mixed-script sample (at most $72$), with similar counts on the stories. This raises the per-sample bound of Proposition~\ref{prop:forge} to $72\cdot2^{-32}\approx1.7\times10^{-8}$. It accepted none of the $1294$ unmarked samples per corpus of Section~\ref{sec:evalfalse}.

\subsubsection{Outer code} The Reed--Solomon outer code raises the untouched threshold to $275$ letters in the books and $293$ in the stories (Table~\ref{tab:threshold}), because its $120$ coded bits need more carriers than the $72$-bit payload. Even well above that threshold it tolerates errors poorly. At $530$ letters, $1.9$ times its threshold in the books, one substitution leaves $0.71$ and $F=4$ pieces leave $0.17$, whereas plain elimination recovers $0.96$ and $0.52$ at the same length. In $5000$ full-rank systems with one flipped equation, the solution was unchanged in $160$ cases, where the flipped row was not a pivot. In the remaining $4840$, at least four of the $15$ symbols were corrupted, more than the three that the outer code corrects, and in every system whose count was recorded the damage reached $13$ to $15$ symbols. Concluding, errors must be corrected on the equations, as decoder~(a) does, before they reach the payload.

\subsection{False accepts and key rotation} \label{sec:evalfalse}
\subsubsection{False accepts}
Table~\ref{tab:false} reports how the extractor treats unmarked text. On mixed-script text none of over $800$ samples per corpus was accepted. The elimination decoder spent $4702$ tag checks on the book samples and $1731$ on the story samples, for $q\,2^{-32}\approx1.1\times10^{-6}$ and $4.0\times10^{-7}$ expected acceptances. In a second set of $450$ such samples, $86$ book samples and $261$ story samples were rejected before any tag check, $82$ and $251$ of them because random variants cancel the votes on shared indices and leave fewer than $72$ usable equations; the tag rejected the rest. Story samples fall below this count more often, since two story sentences hold fewer carriers than two book sentences. Pure-Latin text reads $0$ at every carrier and forms a homogeneous system, which decodes to the all-zero payload whenever it reaches full rank ($363$ of $450$ book samples and $204$ of $450$ story samples). The tag rejects that payload, but each attempt costs a query. Rejecting samples without a non-Latin carrier before decoding removes $5.8$ and $2.38$ tag checks per pure-Latin sample from $q$ in Proposition~\ref{prop:forge}, and the rule never triggered on $8000$ marked excerpts of $150$ to $350$ letters. With it, $0$ of $450$ pure-Latin and $0$ of $844$ mixed-script samples were accepted in both corpora.

\subsubsection{Key rotation}
A verifier that does not know the epoch tries every epoch key. With $\mathcal{E}=16$ epoch keys the tag checks grew by a factor of $15.90$ to $16.05$, close to the expected $16$, and no unmarked sample was accepted. Each of $16$ marks was recovered under its own key, and none of $240$ cross-epoch extractions accepted anything.

\subsection{Splicing} \label{sec:evalsplice}
We concatenated texts marked for two users under one key, with $20$ keys per configuration. With $2000+2000$ letters, $19$ of $20$ book samples were rejected and one was attributed to the first mark, while all $20$ story samples were rejected. With $2000+500$ letters, $14$ of $20$ book samples and $18$ of $20$ story samples were attributed to the longer mark and the others rejected; with $4000+300$ letters, all $20$ were attributed to the longer mark in both corpora. No sample was ever attributed to the shorter mark, and no spliced sample produced an identifier that had not been embedded. In particular, a fragment marked for one user and appended to a longer text of another user did not implicate the first user in any trial.
\begin{table}[t]
\caption{Unmarked samples, tag checks and key rotation.}
\label{tab:false}
\centering
\footnotesize
\setlength{\tabcolsep}{3pt}
\begin{tabular}{lcc}
\toprule
 & Books & LLM stories \\
\midrule
Pure-Latin samples accepted & $0/450$ & $0/450$ \\
\quad full-rank (all-zero) systems & $363/450$ & $204/450$ \\
\quad tag checks/sample, elimination & $5.80$ & $2.38$ \\
\quad tag checks/sample, pre-decode rule & $0.00$ & $0.00$ \\
Mixed-script samples accepted & $0/844$ & $0/844$ \\
Second mixed set: rejected before a tag check & $86/450$ & $261/450$ \\
\quad tag checks/sample, elimination & $5.57$ & $2.05$ \\
\quad tag checks/sample, (a) (max) & $18.07$ ($72$) & $19.84$ ($72$) \\
Marked, $1$ subst., (a) tag checks (max) & $4.70$ ($9$) & $4.80$ ($10$) \\
Pre-decode rule fired on marked text & $0/8000$ & $0/8000$ \\
\midrule
Tag checks, $\mathcal{E}=16$ vs.\ $1$ (Latin/mixed) & $15.90$/$15.95$ & $15.91$/$16.05$ \\
Own-epoch marks recovered & $16/16$ & $16/16$ \\
Cross-epoch extractions accepted & $0/240$ & $0/240$ \\
\bottomrule
\end{tabular}

\end{table}

\subsection{Comparison with a positional codec} \label{sec:evalbaseline}
We compare \name with the positional codec of Rizzo et al~\cite{rizzo}. We do so by applying the same changes to the positional codec of Rizzo et al. and to \name, on the same carriers and documents. The positional codec stores its payload in the first $K$ carriers in reading order, so any deletion in that prefix shifts every later digit.
Table~\ref{tab:baseline} shows the outcome of this comparison. We notice that the positional codec survived a carrier deleted at a random position in $0.95$ of the book trials, close to the $0.96$ probability that the deletion missed the first $K$ carriers. It survived a deletion inside the prefix in less than $2\%$ of the trials and a reordering of sentences in less than $0.5\%$ ($3\%$ on the story). \name recovered the identifier in every trial. 

\begin{table}[t]
\caption{Positional codec against \name on the single documents ($256$ trials per cell).}
\label{tab:baseline}
\centering
\footnotesize
\setlength{\tabcolsep}{3pt}
\begin{tabular}{lcccc}
\toprule
 & \multicolumn{2}{c}{Books} & \multicolumn{2}{c}{LLM stories} \\
\cmidrule(lr){2-3}\cmidrule(lr){4-5}
Change & Positional & \name & Positional & \name \\
\midrule
Clean copy & $1.000$ & $1.000$ & $1.000$ & $1.000$ \\
One carrier deleted, anywhere & $0.953$ & $1.000$ & $0.961$ & $1.000$ \\
One carrier deleted, first $K$ & $0.016$ & $1.000$ & $0.016$ & $1.000$ \\
Sentences reordered & $0.004$ & $1.000$ & $0.031$ & $1.000$ \\
\bottomrule
\end{tabular}

\end{table}

\begin{table}[t]
\caption{Mixed-script check applied word by word.}
\label{tab:detect}
\centering
\footnotesize
\setlength{\tabcolsep}{3pt}
\begin{tabular}{lcc}
\toprule
 & Books & LLM stories \\
\midrule
Unmarked words flagged & $0/475178$ & $0/56658$ \\
Marked words flagged & $0.69$ & $0.65$ \\
$\ge1$ flag in 2 consecutive marked words & $0.91$ & $0.90$ \\
$\ge1$ flag in 3 consecutive marked words & $0.97$ & $0.97$ \\
$\ge1$ flag in 5 consecutive marked words & $0.99$ & $1.00$ \\
$\ge1$ flag in 10 consecutive marked words & $1.00$ & $1.00$ \\
\bottomrule
\end{tabular}

\end{table}
\subsection{Detectability} \label{sec:evaldetect}
Proposition~\ref{prop:region} predicts that a mixed-script test detects a full-weight mark after a few words. To verify this, we applied a mixed-script check following UTS\,\#39~\cite{tr39}, which flags a word that mixes Latin and Cyrillic letters, to every word of the unmarked corpora and of one marked document per corpus as we can see in Table~\ref{tab:detect}. The check flagged none of the unmarked words of the books or the LLM stories, but $0.69$ and $0.65$ of the marked ones. As we can see, there is a $97\%$ chance that 3 consecutive words will contain a flagged word in both corpora.
A marked excerpt long enough to be attributed is thus also long enough to be recognized as marked without the key.

\subsection{Overhead} \label{sec:evaloverhead}
Since \name changes no visible character, we measure the cost in text length and in verification queries. The $72$-bit identifier has its $50\%$ point at $167$ to $178$ letters and is sized for $320$ letters at the design point. A $64$-bit tag for an open verifier adds about $140$ letters, and the asymmetric mode needs $2000$ to $2600$ letters. 
Embedding and extraction are also cheap in time. Typically a harness run that embeds, attacks and extracts the single-document configurations completes, on average, in about $6$ seconds.

\section{Related Work} \label{sec:related}
We group prior work by where the mark resides: in metadata, in the token distribution of a model, in the wording of finished text, or in its format and code points. 

\subsection{Provenance metadata}
Content credentials~\cite{c2pa} bind a signed manifest to a file and are the standard for media. Metadata, however, is often stripped when content is shared across platforms~\cite{nistsynth}, and text copied out of a page carries none. Without a mark, the remaining option is to infer origin from the text itself. Krishna et al.~\cite{krishna} show that paraphrasing evades such detectors and propose retrieval over the provider's own outputs as a defense. Sadasivan et al.~\cite{sadasivan} show that recursive paraphrasing degrades watermark-based, neural, and zero-shot detectors alike. Even when detection works, it can at best establish that a model wrote a text, but not which platform released it to which account.

\subsection{Generation-time watermarks}
A large body of work~\cite{sokwm}, embeds a signal in the token distribution while the model generates. Kirchenbauer et al.~\cite{kirchenbauer} bias sampling toward a keyed ``green list'' of tokens, while SynthID-Text~\cite{synthidtext} deploys tournament sampling in a production system. 
Multi-bit schemes embed a message such as a user identifier in the same distribution. CredID~\cite{credid} attributes through a trusted third party. However, an adversary can learn a watermark from its outputs and then spoof or scrub it~\cite{wmstealing}. While \name and generation-time watermarks have similar goals, they rely on different assumptions. A generation-time watermark requires control over the model's sampling and marks only text that the model produced. \name requires no access to the model and can mark any finished text, and these approaches can be applied in tandem with \name to improve robustness. 

\subsection{Watermarking finished text}
Linguistic steganography and watermarking change the words of a cover text~\cite{lingstego}. Approaches such as synonym substitution~\cite{topkara}, multi-bit marking on features that are resistant to corruption~\cite{yoo}, and neural schemes that rewrite the text via a learned encoder and read it with a learned decoder~\cite{awt,remarkllm} share our deployment model, but they are different from our model in that they change what the reader sees and rely on trained or semantic decoders.

\subsection{Watermarks in format and code points}
Early document watermarking shifted lines and words of rendered documents to trace copies to recipients~\cite{brassil}. Text-only schemes hide data in Unicode space characters~\cite{unispach}, in visually identical code-point substitutions applied to LLM output~\cite{easymark}, or in zero-width characters.
Rizzo et al.~\cite{rizzo} first used homoglyph substitution to watermark text while preserving its appearance, writing a password-derived watermark into confusable letters and whitespace in reading order, and later extended the scheme to protect small excerpts~\cite{rizzofine}. \name differs in three aspects: the identifier is authenticated, the index of each carrier is anchored in its content rather than in its position, and the decoder collects an unordered set of votes, which makes it invariant to reordering and tolerant of deletions.

\subsection{Attacks relying on confusables} 
Homograph attacks were used to register domain names that render like trusted ones~\cite{homograph}, and homoglyph databases built from rendered glyphs show that the confusables of Unicode are a small subset of the pairs that fonts draw alike~\cite{shamfinder}. Invisible and confusable code points were also used to hide logic in source code~\cite{trojansource} and perturb the input of NLP models~\cite{badchars}.

\section{Discussion} \label{sec:discussion}
In essence, \name inherits the limitations of the code-point channel on which it is built. For completeness, in this section we discuss potential threat vectors that could affect \name watermark verification.

\noindent\textbf{Removal.} Folding every carrier to its skeleton erases the identifier without any visible change, and the fold is public. The confusable mappings of UTS\,\#39~\cite{tr39} implement it. Proposition~\ref{prop:keying} shows that keeping the placement secret does not help. For this reason, \name should be considered as a provenance signal for text that is reshared without intent to launder it, and it offers no defense against a user who wants the text to be unattributable. Partial folding acts as erasure, which the code tolerates up to $\kappa$ (see Section~\ref{sec:evalerasure}), so only a deliberate, near-complete fold removes the mark. The same holds for a platform that folds confusables on input. In our evaluation, none of the $18$ platforms we tested was folding the input. However, this reflects their current behavior, which may change over time.

\noindent\textbf{Overtness.} A mixed-script check would flag most marked words. In Proposition~\ref{prop:region} we showed that this cannot be avoided at a useful payload. Users and filters can thus tell that a text is marked. Mixed-script words are also among the signals that homograph defenses look for~\cite{tr39,homograph}, and a spam or phishing filter may treat marked text as suspicious. Search, spell-checking, and screen readers may also handle a Cyrillic letter within a Latin word differently from how they handle its Latin look-alike.

\noindent\textbf{Rewriting.} Paraphrasing, translation, retyping, screenshots, and optical character recognition are all common ways of evading a watermark. While \name and a token-based watermark can be combined on the same response, character-level perturbations will affect both. Homoglyph insertion can disrupt tokenization and hide a token-level watermark from its detector~\cite{charpert}. A platform that layers the two should therefore fold the carriers into the skeleton before running the token-watermark detector.

\section{Conclusions} \label{sec:conclusion}
In this paper, we introduced \name, a text watermarking technique based on keyed homoglyph substitutions. \name works by inserting an authenticated identifier into the choice between Latin characters and their Cyrillic homoglyphs. We showed that \name is robust to copy-paste, partial quotation, sentence reordering, and local edits. In our evaluation, the mark survived copy-paste on all $18$ platforms we tested, and \name reached $50\%$ recovery with roughly $173$ letters on average, which is about one or two sentences of ordinary text. Furthermore, we also derived the capacity of the code-point channel and the minimum length of a markable text.


\bibliographystyle{ACM-Reference-Format}
\bibliography{bibliography}


\begin{thebibliography}{51}


\ifx \showCODEN    \undefined \def \showCODEN     #1{\unskip}     \fi
\ifx \showISBNx    \undefined \def \showISBNx     #1{\unskip}     \fi
\ifx \showISBNxiii \undefined \def \showISBNxiii  #1{\unskip}     \fi
\ifx \showISSN     \undefined \def \showISSN      #1{\unskip}     \fi
\ifx \showLCCN     \undefined \def \showLCCN      #1{\unskip}     \fi
\ifx \shownote     \undefined \def \shownote      #1{#1}          \fi
\ifx \showarticletitle \undefined \def \showarticletitle #1{#1}   \fi
\ifx \showURL      \undefined \def \showURL       {\relax}        \fi
\providecommand\bibfield[2]{#2}
\providecommand\bibinfo[2]{#2}
\providecommand\natexlab[1]{#1}
\providecommand\showeprint[2][]{arXiv:#2}

\bibitem[Abdelnabi and Fritz(2021)]%
        {awt}
\bibfield{author}{\bibinfo{person}{Sahar Abdelnabi} {and} \bibinfo{person}{Mario Fritz}.} \bibinfo{year}{2021}\natexlab{}.
\newblock \showarticletitle{Adversarial Watermarking Transformer: Towards Tracing Text Provenance with Data Hiding}. In \bibinfo{booktitle}{\emph{2021 IEEE Symposium on Security and Privacy (SP)}}. \bibinfo{publisher}{IEEE}, \bibinfo{pages}{121--140}.
\newblock
\href{https://doi.org/10.1109/SP40001.2021.00083}{doi:\nolinkurl{10.1109/SP40001.2021.00083}}


\bibitem[Aumasson et~al\mbox{.}(2013)]%
        {blake2}
\bibfield{author}{\bibinfo{person}{Jean-Philippe Aumasson}, \bibinfo{person}{Samuel Neves}, \bibinfo{person}{Zooko Wilcox-O'Hearn}, {and} \bibinfo{person}{Christian Winnerlein}.} \bibinfo{year}{2013}\natexlab{}.
\newblock \showarticletitle{{BLAKE2}: Simpler, Smaller, Fast as {MD5}}. In \bibinfo{booktitle}{\emph{Applied Cryptography and Network Security (ACNS 2013)}} \emph{(\bibinfo{series}{Lecture Notes in Computer Science}, Vol.~\bibinfo{volume}{7954})}. \bibinfo{publisher}{Springer}, \bibinfo{pages}{119--135}.
\newblock
\href{https://doi.org/10.1007/978-3-642-38980-1_8}{doi:\nolinkurl{10.1007/978-3-642-38980-1_8}}


\bibitem[Bellare et~al\mbox{.}(1996)]%
        {hmac}
\bibfield{author}{\bibinfo{person}{Mihir Bellare}, \bibinfo{person}{Ran Canetti}, {and} \bibinfo{person}{Hugo Krawczyk}.} \bibinfo{year}{1996}\natexlab{}.
\newblock \showarticletitle{Keying Hash Functions for Message Authentication}. In \bibinfo{booktitle}{\emph{Advances in Cryptology -- CRYPTO '96}} \emph{(\bibinfo{series}{Lecture Notes in Computer Science}, Vol.~\bibinfo{volume}{1109})}. \bibinfo{publisher}{Springer}, \bibinfo{pages}{1--15}.
\newblock
\href{https://doi.org/10.1007/3-540-68697-5_1}{doi:\nolinkurl{10.1007/3-540-68697-5_1}}


\bibitem[Bellare and Namprempre(2000)]%
        {etm}
\bibfield{author}{\bibinfo{person}{Mihir Bellare} {and} \bibinfo{person}{Chanathip Namprempre}.} \bibinfo{year}{2000}\natexlab{}.
\newblock \showarticletitle{Authenticated Encryption: Relations Among Notions and Analysis of the Generic Composition Paradigm}. In \bibinfo{booktitle}{\emph{Advances in Cryptology -- ASIACRYPT 2000}} \emph{(\bibinfo{series}{Lecture Notes in Computer Science}, Vol.~\bibinfo{volume}{1976})}. \bibinfo{publisher}{Springer}, \bibinfo{pages}{531--545}.
\newblock
\href{https://doi.org/10.1007/3-540-44448-3_41}{doi:\nolinkurl{10.1007/3-540-44448-3_41}}


\bibitem[Bennett(2004)]%
        {lingstego}
\bibfield{author}{\bibinfo{person}{Krista Bennett}.} \bibinfo{year}{2004}\natexlab{}.
\newblock \bibinfo{booktitle}{\emph{Linguistic Steganography: Survey, Analysis, and Robustness Concerns for Hiding Information in Text}}.
\newblock \bibinfo{type}{CERIAS Tech. Rep.} 2004-13. \bibinfo{institution}{Center for Education and Research in Information Assurance and Security (CERIAS), Purdue University}.
\newblock


\bibitem[Bernstein et~al\mbox{.}(2012)]%
        {eddsa}
\bibfield{author}{\bibinfo{person}{Daniel~J. Bernstein}, \bibinfo{person}{Niels Duif}, \bibinfo{person}{Tanja Lange}, \bibinfo{person}{Peter Schwabe}, {and} \bibinfo{person}{Bo-Yin Yang}.} \bibinfo{year}{2012}\natexlab{}.
\newblock \showarticletitle{High-Speed High-Security Signatures}.
\newblock \bibinfo{journal}{\emph{Journal of Cryptographic Engineering}} \bibinfo{volume}{2}, \bibinfo{number}{2} (\bibinfo{year}{2012}), \bibinfo{pages}{77--89}.
\newblock
\href{https://doi.org/10.1007/s13389-012-0027-1}{doi:\nolinkurl{10.1007/s13389-012-0027-1}}


\bibitem[Boucher and Anderson(2023)]%
        {trojansource}
\bibfield{author}{\bibinfo{person}{Nicholas Boucher} {and} \bibinfo{person}{Ross Anderson}.} \bibinfo{year}{2023}\natexlab{}.
\newblock \showarticletitle{Trojan Source: Invisible Vulnerabilities}. In \bibinfo{booktitle}{\emph{32nd USENIX Security Symposium (USENIX Security 23)}}. \bibinfo{publisher}{USENIX Association}, \bibinfo{pages}{6507--6524}.
\newblock


\bibitem[Boucher et~al\mbox{.}(2022)]%
        {badchars}
\bibfield{author}{\bibinfo{person}{Nicholas Boucher}, \bibinfo{person}{Ilia Shumailov}, \bibinfo{person}{Ross Anderson}, {and} \bibinfo{person}{Nicolas Papernot}.} \bibinfo{year}{2022}\natexlab{}.
\newblock \showarticletitle{Bad Characters: Imperceptible {NLP} Attacks}. In \bibinfo{booktitle}{\emph{2022 IEEE Symposium on Security and Privacy (SP)}}. \bibinfo{publisher}{IEEE}, \bibinfo{pages}{1987--2004}.
\newblock
\href{https://doi.org/10.1109/SP46214.2022.9833641}{doi:\nolinkurl{10.1109/SP46214.2022.9833641}}


\bibitem[Brassil et~al\mbox{.}(1995)]%
        {brassil}
\bibfield{author}{\bibinfo{person}{Jack~T. Brassil}, \bibinfo{person}{Steven Low}, \bibinfo{person}{Nicholas~F. Maxemchuk}, {and} \bibinfo{person}{Lawrence O'Gorman}.} \bibinfo{year}{1995}\natexlab{}.
\newblock \showarticletitle{Electronic Marking and Identification Techniques to Discourage Document Copying}.
\newblock \bibinfo{journal}{\emph{IEEE Journal on Selected Areas in Communications}} \bibinfo{volume}{13}, \bibinfo{number}{8} (\bibinfo{year}{1995}), \bibinfo{pages}{1495--1504}.
\newblock
\href{https://doi.org/10.1109/49.464718}{doi:\nolinkurl{10.1109/49.464718}}


\bibitem[Cachin(2004)]%
        {cachin}
\bibfield{author}{\bibinfo{person}{Christian Cachin}.} \bibinfo{year}{2004}\natexlab{}.
\newblock \showarticletitle{An Information-Theoretic Model for Steganography}.
\newblock \bibinfo{journal}{\emph{Information and Computation}} \bibinfo{volume}{192}, \bibinfo{number}{1} (\bibinfo{year}{2004}), \bibinfo{pages}{41--56}.
\newblock
\href{https://doi.org/10.1016/j.ic.2004.02.003}{doi:\nolinkurl{10.1016/j.ic.2004.02.003}}


\bibitem[{Coalition for Content Provenance and Authenticity}(2023)]%
        {c2pa}
\bibfield{author}{\bibinfo{person}{{Coalition for Content Provenance and Authenticity}}.} \bibinfo{year}{2023}\natexlab{}.
\newblock \bibinfo{booktitle}{\emph{{C2PA} Technical Specification}}.
\newblock \bibinfo{type}{{T}echnical {R}eport}. \bibinfo{institution}{Coalition for Content Provenance and Authenticity (C2PA)}.
\newblock
\urldef\tempurl%
\url{https://c2pa.org/specifications/specifications/1.3/specs/C2PA_Specification.html}
\showURL{%
\tempurl}


\bibitem[Cover and Thomas(2006)]%
        {cover}
\bibfield{author}{\bibinfo{person}{Thomas~M. Cover} {and} \bibinfo{person}{Joy~A. Thomas}.} \bibinfo{year}{2006}\natexlab{}.
\newblock \bibinfo{booktitle}{\emph{Elements of Information Theory} (\bibinfo{edition}{2nd} ed.)}.
\newblock \bibinfo{publisher}{Wiley-Interscience}.
\newblock


\bibitem[Creo and Pudasaini(2025)]%
        {silverspeak}
\bibfield{author}{\bibinfo{person}{Aldan Creo} {and} \bibinfo{person}{Shushanta Pudasaini}.} \bibinfo{year}{2025}\natexlab{}.
\newblock \showarticletitle{{SilverSpeak}: Evading {AI}-Generated Text Detectors using Homoglyphs}. In \bibinfo{booktitle}{\emph{Proceedings of the 1st Workshop on GenAI Content Detection (GenAIDetect)}}. \bibinfo{publisher}{International Committee on Computational Linguistics}, \bibinfo{pages}{1--46}.
\newblock
\urldef\tempurl%
\url{https://arxiv.org/abs/2406.11239}
\showURL{%
\tempurl}


\bibitem[Dathathri et~al\mbox{.}(2024)]%
        {synthidtext}
\bibfield{author}{\bibinfo{person}{Sumanth Dathathri}, \bibinfo{person}{Abigail See}, \bibinfo{person}{Sumedh Ghaisas}, \bibinfo{person}{Po-Sen Huang}, \bibinfo{person}{Rob McAdam}, \bibinfo{person}{Johannes Welbl}, \bibinfo{person}{Vandana Bachani}, \bibinfo{person}{Alex Kaskasoli}, \bibinfo{person}{Robert Stanforth}, \bibinfo{person}{Tatiana Matejovicova}, \bibinfo{person}{Jamie Hayes}, \bibinfo{person}{Nidhi Vyas}, \bibinfo{person}{Majd Al~Merey}, \bibinfo{person}{Jonah Brown-Cohen}, \bibinfo{person}{Rudy Bunel}, \bibinfo{person}{Borja Balle}, \bibinfo{person}{Taylan Cemgil}, \bibinfo{person}{Zahra Ahmed}, \bibinfo{person}{Kitty Stacpoole}, \bibinfo{person}{Ilia Shumailov}, \bibinfo{person}{Ciprian Baetu}, \bibinfo{person}{Sven Gowal}, \bibinfo{person}{Demis Hassabis}, {and} \bibinfo{person}{Pushmeet Kohli}.} \bibinfo{year}{2024}\natexlab{}.
\newblock \showarticletitle{Scalable Watermarking for Identifying Large Language Model Outputs}.
\newblock \bibinfo{journal}{\emph{Nature}} \bibinfo{volume}{634}, \bibinfo{number}{8035} (\bibinfo{year}{2024}), \bibinfo{pages}{818--823}.
\newblock
\href{https://doi.org/10.1038/s41586-024-08025-4}{doi:\nolinkurl{10.1038/s41586-024-08025-4}}


\bibitem[{European Parliament and Council of the European Union}(2024)]%
        {euaiact}
\bibfield{author}{\bibinfo{person}{{European Parliament and Council of the European Union}}.} \bibinfo{year}{2024}\natexlab{}.
\newblock \bibinfo{title}{Regulation ({EU}) 2024/1689 of 13 {June} 2024 Laying Down Harmonised Rules on Artificial Intelligence ({Artificial Intelligence Act})}.
\newblock \bibinfo{howpublished}{Official Journal of the European Union, L~2024/1689}.
\newblock
\urldef\tempurl%
\url{https://eur-lex.europa.eu/eli/reg/2024/1689/oj}
\showURL{%
\tempurl}
\newblock
\shownote{{Art.}~50(2)}.


\bibitem[Ferrara et~al\mbox{.}(2016)]%
        {ferrara}
\bibfield{author}{\bibinfo{person}{Emilio Ferrara}, \bibinfo{person}{Onur Varol}, \bibinfo{person}{Clayton Davis}, \bibinfo{person}{Filippo Menczer}, {and} \bibinfo{person}{Alessandro Flammini}.} \bibinfo{year}{2016}\natexlab{}.
\newblock \showarticletitle{The Rise of Social Bots}.
\newblock \bibinfo{journal}{\emph{Commun. ACM}} \bibinfo{volume}{59}, \bibinfo{number}{7} (\bibinfo{year}{2016}), \bibinfo{pages}{96--104}.
\newblock
\href{https://doi.org/10.1145/2818717}{doi:\nolinkurl{10.1145/2818717}}


\bibitem[Gabrilovich and Gontmakher(2002)]%
        {homograph}
\bibfield{author}{\bibinfo{person}{Evgeniy Gabrilovich} {and} \bibinfo{person}{Alex Gontmakher}.} \bibinfo{year}{2002}\natexlab{}.
\newblock \showarticletitle{The Homograph Attack}.
\newblock \bibinfo{journal}{\emph{Commun. ACM}} \bibinfo{volume}{45}, \bibinfo{number}{2} (\bibinfo{year}{2002}), \bibinfo{pages}{128}.
\newblock
\href{https://doi.org/10.1145/503124.503156}{doi:\nolinkurl{10.1145/503124.503156}}


\bibitem[Goldstein et~al\mbox{.}(2023)]%
        {goldstein}
\bibfield{author}{\bibinfo{person}{Josh~A. Goldstein}, \bibinfo{person}{Girish Sastry}, \bibinfo{person}{Micah Musser}, \bibinfo{person}{Ren{\'e}e DiResta}, \bibinfo{person}{Mark Gentzel}, {and} \bibinfo{person}{Katerina Sedova}.} \bibinfo{year}{2023}\natexlab{}.
\newblock \showarticletitle{Generative Language Models and Automated Influence Operations: Emerging Threats and Potential Mitigations}.
\newblock \bibinfo{journal}{\emph{arXiv preprint arXiv:2301.04246}} (\bibinfo{year}{2023}).
\newblock
\urldef\tempurl%
\url{https://arxiv.org/abs/2301.04246}
\showURL{%
\tempurl}


\bibitem[Goldwasser et~al\mbox{.}(1988)]%
        {gmr}
\bibfield{author}{\bibinfo{person}{Shafi Goldwasser}, \bibinfo{person}{Silvio Micali}, {and} \bibinfo{person}{Ronald~L. Rivest}.} \bibinfo{year}{1988}\natexlab{}.
\newblock \showarticletitle{A Digital Signature Scheme Secure Against Adaptive Chosen-Message Attacks}.
\newblock \bibinfo{journal}{\emph{SIAM J. Comput.}} \bibinfo{volume}{17}, \bibinfo{number}{2} (\bibinfo{year}{1988}), \bibinfo{pages}{281--308}.
\newblock
\href{https://doi.org/10.1137/0217017}{doi:\nolinkurl{10.1137/0217017}}


\bibitem[Gueron et~al\mbox{.}(2019)]%
        {rfc8452}
\bibfield{author}{\bibinfo{person}{Shay Gueron}, \bibinfo{person}{Adam Langley}, {and} \bibinfo{person}{Yehuda Lindell}.} \bibinfo{year}{2019}\natexlab{}.
\newblock \bibinfo{booktitle}{\emph{{AES-GCM-SIV}: Nonce Misuse-Resistant Authenticated Encryption}}.
\newblock \bibinfo{type}{RFC} 8452. \bibinfo{institution}{RFC Editor}.
\newblock
\href{https://doi.org/10.17487/RFC8452}{doi:\nolinkurl{10.17487/RFC8452}}


\bibitem[Hopper et~al\mbox{.}(2002)]%
        {hla}
\bibfield{author}{\bibinfo{person}{Nicholas~J. Hopper}, \bibinfo{person}{John Langford}, {and} \bibinfo{person}{Luis von Ahn}.} \bibinfo{year}{2002}\natexlab{}.
\newblock \showarticletitle{Provably Secure Steganography}. In \bibinfo{booktitle}{\emph{Advances in Cryptology -- CRYPTO 2002}} \emph{(\bibinfo{series}{Lecture Notes in Computer Science}, Vol.~\bibinfo{volume}{2442})}. \bibinfo{publisher}{Springer}, \bibinfo{pages}{77--92}.
\newblock
\href{https://doi.org/10.1007/3-540-45708-9_6}{doi:\nolinkurl{10.1007/3-540-45708-9_6}}


\bibitem[Jiang et~al\mbox{.}(2024)]%
        {credid}
\bibfield{author}{\bibinfo{person}{Haoyu Jiang}, \bibinfo{person}{Xuhong Wang}, \bibinfo{person}{Ping Yi}, \bibinfo{person}{Shanzhe Lei}, {and} \bibinfo{person}{Yilun Lin}.} \bibinfo{year}{2024}\natexlab{}.
\newblock \showarticletitle{{CredID}: Credible Multi-Bit Watermark for Large Language Models Identification}.
\newblock \bibinfo{journal}{\emph{arXiv preprint arXiv:2412.03107}} (\bibinfo{year}{2024}).
\newblock
\urldef\tempurl%
\url{https://arxiv.org/abs/2412.03107}
\showURL{%
\tempurl}


\bibitem[Jovanovi{\'c} et~al\mbox{.}(2024)]%
        {wmstealing}
\bibfield{author}{\bibinfo{person}{Nikola Jovanovi{\'c}}, \bibinfo{person}{Robin Staab}, {and} \bibinfo{person}{Martin Vechev}.} \bibinfo{year}{2024}\natexlab{}.
\newblock \showarticletitle{Watermark Stealing in Large Language Models}. In \bibinfo{booktitle}{\emph{Proceedings of the 41st International Conference on Machine Learning (ICML)}} \emph{(\bibinfo{series}{Proceedings of Machine Learning Research}, Vol.~\bibinfo{volume}{235})}. \bibinfo{publisher}{PMLR}, \bibinfo{pages}{22570--22593}.
\newblock


\bibitem[Kirchenbauer et~al\mbox{.}(2023)]%
        {kirchenbauer}
\bibfield{author}{\bibinfo{person}{John Kirchenbauer}, \bibinfo{person}{Jonas Geiping}, \bibinfo{person}{Yuxin Wen}, \bibinfo{person}{Jonathan Katz}, \bibinfo{person}{Ian Miers}, {and} \bibinfo{person}{Tom Goldstein}.} \bibinfo{year}{2023}\natexlab{}.
\newblock \showarticletitle{A Watermark for Large Language Models}. In \bibinfo{booktitle}{\emph{Proceedings of the 40th International Conference on Machine Learning (ICML)}} \emph{(\bibinfo{series}{Proceedings of Machine Learning Research}, Vol.~\bibinfo{volume}{202})}. \bibinfo{publisher}{PMLR}, \bibinfo{pages}{17061--17084}.
\newblock


\bibitem[Krishna et~al\mbox{.}(2023)]%
        {krishna}
\bibfield{author}{\bibinfo{person}{Kalpesh Krishna}, \bibinfo{person}{Yixiao Song}, \bibinfo{person}{Marzena Karpinska}, \bibinfo{person}{John Wieting}, {and} \bibinfo{person}{Mohit Iyyer}.} \bibinfo{year}{2023}\natexlab{}.
\newblock \showarticletitle{Paraphrasing Evades Detectors of {AI}-Generated Text, but Retrieval is an Effective Defense}. In \bibinfo{booktitle}{\emph{Advances in Neural Information Processing Systems (NeurIPS)}}, Vol.~\bibinfo{volume}{36}.
\newblock


\bibitem[Luby(2002)]%
        {ltcodes}
\bibfield{author}{\bibinfo{person}{Michael Luby}.} \bibinfo{year}{2002}\natexlab{}.
\newblock \showarticletitle{{LT} Codes}. In \bibinfo{booktitle}{\emph{Proceedings of the 43rd Annual IEEE Symposium on Foundations of Computer Science (FOCS)}}. \bibinfo{publisher}{IEEE}, \bibinfo{pages}{271--280}.
\newblock
\href{https://doi.org/10.1109/SFCS.2002.1181950}{doi:\nolinkurl{10.1109/SFCS.2002.1181950}}


\bibitem[MacKay(2005)]%
        {mackay}
\bibfield{author}{\bibinfo{person}{David J.~C. MacKay}.} \bibinfo{year}{2005}\natexlab{}.
\newblock \showarticletitle{Fountain Codes}.
\newblock \bibinfo{journal}{\emph{IEE Proceedings -- Communications}} \bibinfo{volume}{152}, \bibinfo{number}{6} (\bibinfo{year}{2005}), \bibinfo{pages}{1062--1068}.
\newblock


\bibitem[{National Institute of Standards and Technology}(2024)]%
        {nistsynth}
\bibfield{author}{\bibinfo{person}{{National Institute of Standards and Technology}}.} \bibinfo{year}{2024}\natexlab{}.
\newblock \bibinfo{booktitle}{\emph{Reducing Risks Posed by Synthetic Content: An Overview of Technical Approaches to Digital Content Transparency}}.
\newblock \bibinfo{type}{NIST AI} 100-4. \bibinfo{institution}{National Institute of Standards and Technology}.
\newblock
\href{https://doi.org/10.6028/NIST.AI.100-4}{doi:\nolinkurl{10.6028/NIST.AI.100-4}}


\bibitem[Norvig(2013)]%
        {norvig}
\bibfield{author}{\bibinfo{person}{Peter Norvig}.} \bibinfo{year}{2013}\natexlab{}.
\newblock \bibinfo{title}{English Letter Frequency Counts: {Mayzner} Revisited}.
\newblock
\urldef\tempurl%
\url{https://norvig.com/mayzner.html}
\showURL{%
\tempurl}


\bibitem[Pajola and Conti(2021)]%
        {zew}
\bibfield{author}{\bibinfo{person}{Luca Pajola} {and} \bibinfo{person}{Mauro Conti}.} \bibinfo{year}{2021}\natexlab{}.
\newblock \showarticletitle{Fall of Giants: How Popular Text-Based {MLaaS} Fall against a Simple Evasion Attack}. In \bibinfo{booktitle}{\emph{Proceedings of the 6th IEEE European Symposium on Security and Privacy (EuroS\&P)}}. \bibinfo{pages}{198--211}.
\newblock
\href{https://doi.org/10.1109/EuroSP51992.2021.00023}{doi:\nolinkurl{10.1109/EuroSP51992.2021.00023}}


\bibitem[Por et~al\mbox{.}(2012)]%
        {unispach}
\bibfield{author}{\bibinfo{person}{Lip~Yee Por}, \bibinfo{person}{KokSheik Wong}, {and} \bibinfo{person}{Kok~Onn Chee}.} \bibinfo{year}{2012}\natexlab{}.
\newblock \showarticletitle{{UniSpaCh}: A Text-Based Data Hiding Method Using {Unicode} Space Characters}.
\newblock \bibinfo{journal}{\emph{Journal of Systems and Software}} \bibinfo{volume}{85}, \bibinfo{number}{5} (\bibinfo{year}{2012}), \bibinfo{pages}{1075--1082}.
\newblock
\href{https://doi.org/10.1016/j.jss.2011.12.023}{doi:\nolinkurl{10.1016/j.jss.2011.12.023}}


\bibitem[{Project Gutenberg}({[n.\,d.]})]%
        {gutenberg}
\bibfield{author}{\bibinfo{person}{{Project Gutenberg}}.} \bibinfo{year}{[n.\,d.]}\natexlab{}.
\newblock \bibinfo{title}{Ebooks \#1404 (The Federalist Papers), \#1342 (Pride and Prejudice), and \#1228 (On the Origin of Species)}.
\newblock
\urldef\tempurl%
\url{https://www.gutenberg.org}
\showURL{%
\tempurl}


\bibitem[Qu et~al\mbox{.}(2025)]%
        {qu}
\bibfield{author}{\bibinfo{person}{Wenjie Qu}, \bibinfo{person}{Wengrui Zheng}, \bibinfo{person}{Tianyang Tao}, \bibinfo{person}{Dong Yin}, \bibinfo{person}{Yanze Jiang}, \bibinfo{person}{Zhihua Tian}, \bibinfo{person}{Wei Zou}, \bibinfo{person}{Jinyuan Jia}, {and} \bibinfo{person}{Jiaheng Zhang}.} \bibinfo{year}{2025}\natexlab{}.
\newblock \showarticletitle{Provably Robust Multi-bit Watermarking for {AI}-generated Text}. In \bibinfo{booktitle}{\emph{34th USENIX Security Symposium (USENIX Security 25)}}. \bibinfo{publisher}{USENIX Association}, \bibinfo{pages}{201--220}.
\newblock
\newblock
\shownote{Extended version: arXiv:2401.16820}.


\bibitem[Reed and Solomon(1960)]%
        {reedsolomon}
\bibfield{author}{\bibinfo{person}{Irving~S. Reed} {and} \bibinfo{person}{Gustave Solomon}.} \bibinfo{year}{1960}\natexlab{}.
\newblock \showarticletitle{Polynomial Codes over Certain Finite Fields}.
\newblock \bibinfo{journal}{\emph{J. Soc. Indust. Appl. Math.}} \bibinfo{volume}{8}, \bibinfo{number}{2} (\bibinfo{year}{1960}), \bibinfo{pages}{300--304}.
\newblock
\href{https://doi.org/10.1137/0108018}{doi:\nolinkurl{10.1137/0108018}}


\bibitem[Rizzo et~al\mbox{.}(2016)]%
        {rizzo}
\bibfield{author}{\bibinfo{person}{Stefano~Giovanni Rizzo}, \bibinfo{person}{Flavio Bertini}, {and} \bibinfo{person}{Danilo Montesi}.} \bibinfo{year}{2016}\natexlab{}.
\newblock \showarticletitle{Content-Preserving Text Watermarking through {Unicode} Homoglyph Substitution}. In \bibinfo{booktitle}{\emph{Proceedings of the 20th International Database Engineering \& Applications Symposium (IDEAS)}}. \bibinfo{publisher}{ACM}, \bibinfo{pages}{97--104}.
\newblock
\href{https://doi.org/10.1145/2938503.2938510}{doi:\nolinkurl{10.1145/2938503.2938510}}


\bibitem[Rizzo et~al\mbox{.}(2019)]%
        {rizzofine}
\bibfield{author}{\bibinfo{person}{Stefano~Giovanni Rizzo}, \bibinfo{person}{Flavio Bertini}, {and} \bibinfo{person}{Danilo Montesi}.} \bibinfo{year}{2019}\natexlab{}.
\newblock \showarticletitle{Fine-Grain Watermarking for Intellectual Property Protection}.
\newblock \bibinfo{journal}{\emph{EURASIP Journal on Information Security}} \bibinfo{volume}{2019}, \bibinfo{number}{1} (\bibinfo{year}{2019}).
\newblock
\href{https://doi.org/10.1186/s13635-019-0094-2}{doi:\nolinkurl{10.1186/s13635-019-0094-2}}
\newblock
\shownote{{Art.} no.~10}.


\bibitem[Rogaway and Shrimpton(2006)]%
        {siv}
\bibfield{author}{\bibinfo{person}{Phillip Rogaway} {and} \bibinfo{person}{Thomas Shrimpton}.} \bibinfo{year}{2006}\natexlab{}.
\newblock \showarticletitle{A Provable-Security Treatment of the Key-Wrap Problem}. In \bibinfo{booktitle}{\emph{Advances in Cryptology -- EUROCRYPT 2006}} \emph{(\bibinfo{series}{Lecture Notes in Computer Science}, Vol.~\bibinfo{volume}{4004})}. \bibinfo{publisher}{Springer}, \bibinfo{pages}{373--390}.
\newblock
\href{https://doi.org/10.1007/11761679_23}{doi:\nolinkurl{10.1007/11761679_23}}


\bibitem[Sadasivan et~al\mbox{.}(2025)]%
        {sadasivan}
\bibfield{author}{\bibinfo{person}{Vinu~Sankar Sadasivan}, \bibinfo{person}{Aounon Kumar}, \bibinfo{person}{Sriram Balasubramanian}, \bibinfo{person}{Wenxiao Wang}, {and} \bibinfo{person}{Soheil Feizi}.} \bibinfo{year}{2025}\natexlab{}.
\newblock \showarticletitle{Can {AI}-Generated Text be Reliably Detected? {Stress} Testing {AI} Text Detectors Under Various Attacks}.
\newblock \bibinfo{journal}{\emph{Transactions on Machine Learning Research}} (\bibinfo{year}{2025}).
\newblock
\urldef\tempurl%
\url{https://openreview.net/forum?id=OOgsAZdFOt}
\showURL{%
\tempurl}


\bibitem[Sato et~al\mbox{.}(2023)]%
        {easymark}
\bibfield{author}{\bibinfo{person}{Ryoma Sato}, \bibinfo{person}{Yuki Takezawa}, \bibinfo{person}{Han Bao}, \bibinfo{person}{Kenta Niwa}, {and} \bibinfo{person}{Makoto Yamada}.} \bibinfo{year}{2023}\natexlab{}.
\newblock \showarticletitle{Embarrassingly Simple Text Watermarks}.
\newblock \bibinfo{journal}{\emph{arXiv preprint arXiv:2310.08920}} (\bibinfo{year}{2023}).
\newblock
\urldef\tempurl%
\url{https://arxiv.org/abs/2310.08920}
\showURL{%
\tempurl}


\bibitem[Shokrollahi(2006)]%
        {raptor}
\bibfield{author}{\bibinfo{person}{Amin Shokrollahi}.} \bibinfo{year}{2006}\natexlab{}.
\newblock \showarticletitle{Raptor Codes}.
\newblock \bibinfo{journal}{\emph{IEEE Transactions on Information Theory}} \bibinfo{volume}{52}, \bibinfo{number}{6} (\bibinfo{year}{2006}), \bibinfo{pages}{2551--2567}.
\newblock
\href{https://doi.org/10.1109/TIT.2006.874390}{doi:\nolinkurl{10.1109/TIT.2006.874390}}


\bibitem[Suzuki et~al\mbox{.}(2019)]%
        {shamfinder}
\bibfield{author}{\bibinfo{person}{Hiroaki Suzuki}, \bibinfo{person}{Daiki Chiba}, \bibinfo{person}{Yoshiro Yoneya}, \bibinfo{person}{Tatsuya Mori}, {and} \bibinfo{person}{Shigeki Goto}.} \bibinfo{year}{2019}\natexlab{}.
\newblock \showarticletitle{{ShamFinder}: An Automated Framework for Detecting {IDN} Homographs}. In \bibinfo{booktitle}{\emph{Proceedings of the Internet Measurement Conference (IMC)}}. \bibinfo{publisher}{ACM}, \bibinfo{pages}{449--462}.
\newblock
\href{https://doi.org/10.1145/3355369.3355587}{doi:\nolinkurl{10.1145/3355369.3355587}}


\bibitem[Taleby~Ahvanooey et~al\mbox{.}(2018)]%
        {aitsteg}
\bibfield{author}{\bibinfo{person}{Milad Taleby~Ahvanooey}, \bibinfo{person}{Qianmu Li}, \bibinfo{person}{Jun Hou}, \bibinfo{person}{Hassan Dana~Mazraeh}, {and} \bibinfo{person}{Jing Zhang}.} \bibinfo{year}{2018}\natexlab{}.
\newblock \showarticletitle{{AITSteg}: An Innovative Text Steganography Technique for Hidden Transmission of Text Message via Social Media}.
\newblock \bibinfo{journal}{\emph{IEEE Access}}  \bibinfo{volume}{6} (\bibinfo{year}{2018}), \bibinfo{pages}{65981--65995}.
\newblock
\href{https://doi.org/10.1109/ACCESS.2018.2866063}{doi:\nolinkurl{10.1109/ACCESS.2018.2866063}}


\bibitem[Topkara et~al\mbox{.}(2006)]%
        {topkara}
\bibfield{author}{\bibinfo{person}{Umut Topkara}, \bibinfo{person}{Mercan Topkara}, {and} \bibinfo{person}{Mikhail~J. Atallah}.} \bibinfo{year}{2006}\natexlab{}.
\newblock \showarticletitle{The Hiding Virtues of Ambiguity: Quantifiably Resilient Watermarking of Natural Language Text through Synonym Substitutions}. In \bibinfo{booktitle}{\emph{Proceedings of the 8th Workshop on Multimedia and Security (MM\&Sec)}}. \bibinfo{publisher}{ACM}, \bibinfo{pages}{164--174}.
\newblock
\href{https://doi.org/10.1145/1161366.1161397}{doi:\nolinkurl{10.1145/1161366.1161397}}


\bibitem[{Unicode Consortium}(2023a)]%
        {uax29}
\bibfield{author}{\bibinfo{person}{{Unicode Consortium}}.} \bibinfo{year}{2023}\natexlab{a}.
\newblock \bibinfo{booktitle}{\emph{Unicode Standard Annex \#29: Unicode Text Segmentation}}.
\newblock \bibinfo{type}{UAX}~29. \bibinfo{institution}{Unicode Consortium}.
\newblock
\urldef\tempurl%
\url{https://www.unicode.org/reports/tr29/}
\showURL{%
\tempurl}


\bibitem[{Unicode Consortium}(2023b)]%
        {tr39}
\bibfield{author}{\bibinfo{person}{{Unicode Consortium}}.} \bibinfo{year}{2023}\natexlab{b}.
\newblock \bibinfo{booktitle}{\emph{Unicode Technical Standard \#39: Unicode Security Mechanisms}}.
\newblock \bibinfo{type}{UTS}~39. \bibinfo{institution}{Unicode Consortium}.
\newblock
\urldef\tempurl%
\url{https://www.unicode.org/reports/tr39/}
\showURL{%
\tempurl}


\bibitem[Yang and Menczer(2024)]%
        {aibotnet}
\bibfield{author}{\bibinfo{person}{Kaicheng Yang} {and} \bibinfo{person}{Filippo Menczer}.} \bibinfo{year}{2024}\natexlab{}.
\newblock \showarticletitle{Anatomy of an {AI}-Powered Malicious Social Botnet}.
\newblock \bibinfo{journal}{\emph{Journal of Quantitative Description: Digital Media}}  \bibinfo{volume}{4} (\bibinfo{year}{2024}).
\newblock
\href{https://doi.org/10.51685/jqd.2024.icwsm.7}{doi:\nolinkurl{10.51685/jqd.2024.icwsm.7}}


\bibitem[Yoo et~al\mbox{.}(2023)]%
        {yoo}
\bibfield{author}{\bibinfo{person}{KiYoon Yoo}, \bibinfo{person}{Wonhyuk Ahn}, \bibinfo{person}{Jiho Jang}, {and} \bibinfo{person}{Nojun Kwak}.} \bibinfo{year}{2023}\natexlab{}.
\newblock \showarticletitle{Robust Multi-bit Natural Language Watermarking through Invariant Features}. In \bibinfo{booktitle}{\emph{Proceedings of the 61st Annual Meeting of the Association for Computational Linguistics (Volume 1: Long Papers)}}. \bibinfo{publisher}{Association for Computational Linguistics}, \bibinfo{pages}{2092--2115}.
\newblock
\href{https://doi.org/10.18653/v1/2023.acl-long.117}{doi:\nolinkurl{10.18653/v1/2023.acl-long.117}}


\bibitem[Yoo et~al\mbox{.}(2024)]%
        {mpac}
\bibfield{author}{\bibinfo{person}{KiYoon Yoo}, \bibinfo{person}{Wonhyuk Ahn}, {and} \bibinfo{person}{Nojun Kwak}.} \bibinfo{year}{2024}\natexlab{}.
\newblock \showarticletitle{Advancing Beyond Identification: Multi-bit Watermark for Large Language Models}. In \bibinfo{booktitle}{\emph{Proceedings of the 2024 Conference of the North American Chapter of the Association for Computational Linguistics: Human Language Technologies (Volume 1: Long Papers)}}. \bibinfo{publisher}{Association for Computational Linguistics}, \bibinfo{pages}{4031--4055}.
\newblock
\href{https://doi.org/10.18653/v1/2024.naacl-long.224}{doi:\nolinkurl{10.18653/v1/2024.naacl-long.224}}


\bibitem[Zhang et~al\mbox{.}(2024)]%
        {remarkllm}
\bibfield{author}{\bibinfo{person}{Ruisi Zhang}, \bibinfo{person}{Shehzeen~Samarah Hussain}, \bibinfo{person}{Paarth Neekhara}, {and} \bibinfo{person}{Farinaz Koushanfar}.} \bibinfo{year}{2024}\natexlab{}.
\newblock \showarticletitle{{REMARK-LLM}: A Robust and Efficient Watermarking Framework for Generative Large Language Models}. In \bibinfo{booktitle}{\emph{Proceedings of the 33rd USENIX Security Symposium}}. \bibinfo{pages}{1813--1830}.
\newblock


\bibitem[Zhang et~al\mbox{.}(2026)]%
        {charpert}
\bibfield{author}{\bibinfo{person}{Zhaoxi Zhang}, \bibinfo{person}{Xiaomei Zhang}, \bibinfo{person}{Yanjun Zhang}, \bibinfo{person}{He Zhang}, \bibinfo{person}{Shirui Pan}, \bibinfo{person}{Bo Liu}, \bibinfo{person}{Asif~Qumer Gill}, {and} \bibinfo{person}{Leo~Yu Zhang}.} \bibinfo{year}{2026}\natexlab{}.
\newblock \showarticletitle{Character-Level Perturbations Disrupt {LLM} Watermarks}. In \bibinfo{booktitle}{\emph{Network and Distributed System Security Symposium (NDSS)}}. \bibinfo{publisher}{Internet Society}.
\newblock
\href{https://doi.org/10.14722/ndss.2026.230138}{doi:\nolinkurl{10.14722/ndss.2026.230138}}


\bibitem[Zhao et~al\mbox{.}(2025)]%
        {sokwm}
\bibfield{author}{\bibinfo{person}{Xuandong Zhao}, \bibinfo{person}{Sam Gunn}, \bibinfo{person}{Miranda Christ}, \bibinfo{person}{Jaiden Fairoze}, \bibinfo{person}{Andres Fabrega}, \bibinfo{person}{Nicholas Carlini}, \bibinfo{person}{Sanjam Garg}, \bibinfo{person}{Sanghyun Hong}, \bibinfo{person}{Milad Nasr}, \bibinfo{person}{Florian Tram{\`e}r}, \bibinfo{person}{Somesh Jha}, \bibinfo{person}{Lei Li}, \bibinfo{person}{Yu-Xiang Wang}, {and} \bibinfo{person}{Dawn Song}.} \bibinfo{year}{2025}\natexlab{}.
\newblock \showarticletitle{{SoK}: Watermarking for {AI}-Generated Content}. In \bibinfo{booktitle}{\emph{2025 IEEE Symposium on Security and Privacy (SP)}}. \bibinfo{publisher}{IEEE}, \bibinfo{pages}{2621--2639}.
\newblock
\href{https://doi.org/10.1109/SP61157.2025.00178}{doi:\nolinkurl{10.1109/SP61157.2025.00178}}


\end{thebibliography}

\appendix
\section{Proofs of the Propositions}
\label{app:proofs}

\begin{proof}[Proof of Proposition~\ref{prop:capacity}]
Converse. The extractor's output is a function of the received text, which reveals the surviving carriers. An erased carrier is statistically independent of its embedded symbol. By the data-processing inequality~\cite{cover}, $I(m;\tilde T)\le\sum_{i\ \text{survives}}\log_2 v_i$, so a decoder cannot output more information about $m$ than its input contains. The bound holds pointwise, for every realization of the erasure pattern and every $L$, without any asymptotic argument. Since $(1-p_e)R$ is the mean of the right-hand side, Equation~\eqref{eq:capacity} is its expected value, which is used to size a text for a target $p_e$. For a single sample, the pointwise bound applies using that sample's survivor count.
Achievability. A rateless code~\cite{ltcodes,raptor} recovers the $K$ payload bits from any $(1+\varepsilon)K$ received coded bits. Recovery depends only on how many carriers survive, so decoding succeeds once $(1-p_e)R\ge(1+\varepsilon)K$, and the bound also holds for correlated losses such as truncation.
\end{proof}

\begin{proof}[Proof of Proposition~\ref{prop:keying}]
Carrier positions are determined by the public alphabet (Section~\ref{sec:carriers}) and do not depend on $k$. Each carrier holds a coded bit, and the adversary need not distinguish between used and unused positions. The current bit of a carrier is visible in its code point, and an adversary without $k$ can flip it, or fold it to its Latin skeleton without any visible change. $k$ adds the index that each carrier addresses. Since the code is symbol-symmetric, failure depends only on the number of corrupted carriers. Let $\mathcal{B}_0$ corrupt the same number of carriers as $\mathcal{B}_k$ within the same budget; recovery fails with the same probability.
\end{proof}
\section{Supplementary Tables and Figures}
\label{app:extra}

\begin{table}[H]
\caption{Tolerance and cost of each structural edit.}
\label{tab:structural}
\centering
\scriptsize
\setlength{\tabcolsep}{2pt}
\begin{tabular}{@{}lcc@{}}
\toprule
Edit & Tolerated & Cost \\
\midrule
Reordering of sentences & unbounded & none \\
Deletion of sentences & $b\nu(L_S)\ge(1+\varepsilon)K$ & length only \\
Cuts at sentence bounds & unbounded $F$ & none \\
Cuts inside sentences, outer code & $F\le\theta L_S/2W_{\mathrm{ctx}}\alpha_c$ & $2W_{\mathrm{ctx}}\alpha_c$/fragment \\
\midrule
Bare threshold, books & $167$--$168$ letters & no reserve \\
Bare threshold, LLM stories & $178$ letters & no reserve \\
\bottomrule
\end{tabular}
\end{table}

\begin{table}[H]
\caption{Recovery of $240$-character excerpts with surrounding whitespace, as read and after trimming ($40$ documents, $640$ trials per cell).}
\label{tab:whitespace}
\centering
\footnotesize
\setlength{\tabcolsep}{3pt}
\begin{tabular}{lcccc}
\toprule
 & \multicolumn{2}{c}{Books} & \multicolumn{2}{c}{LLM stories} \\
\cmidrule(lr){2-3}\cmidrule(lr){4-5}
$240$-character excerpt & As read & Trimmed & As read & Trimmed \\
\midrule
No surrounding whitespace & $0.97$ & $0.97$ & $0.81$ & $0.81$ \\
Newline before & $0.89$ & $0.97$ & $0.75$ & $0.81$ \\
Newline after & $0.89$ & $0.97$ & $0.74$ & $0.81$ \\
Two spaces at both ends & $0.81$ & $0.97$ & $0.69$ & $0.81$ \\
\bottomrule
\end{tabular}

\end{table}

\begin{table}[H]
\caption{Pure erasures at the $320$-letter design point ($40$ documents, $8$ keys, $640$ trials per cell).}
\label{tab:erasure}
\centering
\footnotesize
\setlength{\tabcolsep}{3.4pt}
\begin{tabular}{lcccccc}
\toprule
Erased fraction & $0.30$ & $0.40$ & $0.45$ & $0.50$ & $0.55$ & $0.60$ \\
\midrule
Books: recovered & $1.00$ & $1.00$ & $0.94$ & $0.45$ & $0.03$ & $0.00$ \\
\quad distinct indices left & $101$ & $87$ & $80$ & $73$ & $66$ & $59$ \\
LLM stories: recovered & $1.00$ & $0.95$ & $0.69$ & $0.12$ & $0.00$ & $0.00$ \\
\quad distinct indices left & $95$ & $82$ & $75$ & $68$ & $62$ & $55$ \\
\bottomrule
\end{tabular}

\end{table}

\begin{figure}[H]
\centering
\includegraphics[width=0.8\columnwidth]{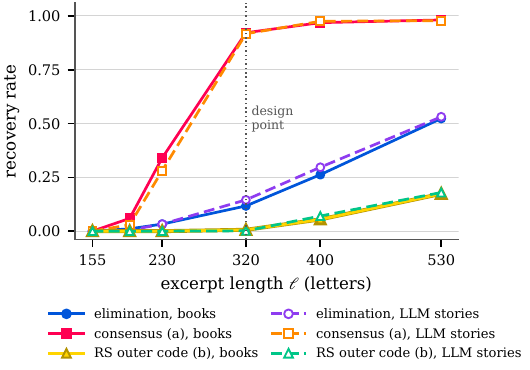}
\caption{Recovery of $\ell$-letter excerpts cut into $F=4$ equal pieces at character offsets, shuffled and rejoined, by decoder and corpus. The dotted line marks the $320$-letter design point.}
\label{fig:deccut}
\end{figure}

\end{document}